\documentclass[final,2p,times]{elsarticle}

\usepackage{epsfig}

\usepackage{algorithm,algpseudocode}

\usepackage{amssymb}
\usepackage{amsthm}

\newcounter{algsubstate}
\renewcommand{\thealgsubstate}{\alph{algsubstate}}
\newenvironment{algsubstates}
  {\setcounter{algsubstate}{0}%
   \renewcommand{\State}{%
     \stepcounter{algsubstate}%
     \Statex {\footnotesize\thealgsubstate:}\space}}
  {}

\usepackage{geometry}                
\usepackage[parfill]{parskip}    
\usepackage{amssymb}
\usepackage{comment}
\usepackage{graphicx,caption}
\usepackage{epstopdf}
\usepackage{hyperref}
\usepackage{bigints}
\usepackage{natbib}
\hypersetup{
    colorlinks = true,
    urlcolor   = blue,
    citecolor  = black,
}
\newtheorem{theorem}{Theorem}

\newtheorem{definition}{Definition}
\theoremstyle{remark}

\begin{document}

\begin{frontmatter}



\title{Material Barriers to the Diffusive Magnetic Flux in Magnetohydrodynamics}

\author[1]{A. Encinas-Bartos}
\author[1]{B. Kaszás}
\author[2]{S. Servidio}

\author[1]{G. Haller \corref{cor1}}
\ead{georgehaller@ethz.ch}

 \cortext[cor1]{Corresponding author}

 \affiliation[1]{organization={Institute for Mechanical Systems, ETH Zurich}, 
                 city={Zurich}, 
                 country={Switzerland}}
 \affiliation[2]{organization = {Dipartimento di Fisica, Università della Calabria}, city={Cosenza}, country = {Italy}}


\begin{abstract}
Recent work has identified objective (frame-indifferent) material barriers that inhibit the transport of dynamically active vectorial quantities (such as linear momentum, angular momentum and vorticity) in Navier-Stokes flows. In magnetohydrodynamics (MHD), a similar setting arises: the magnetic field vector impacts the evolution of the velocity field through the Lorentz force and hence is a dynamically active vector field. Here, we extend the theory of active material barriers from Navier-Stokes flows to MHD flows in order to locate frame-indifferent barriers that minimize the diffusive magnetic flux in turbulent two-dimensional and three-dimensional MHD flows. From this approach we obtain an algorithm for the automated extraction of such barriers from MHD turbulence data. Our findings suggest that the identified barriers inhibit magnetic diffusion, separate electric current sheets and organise the transport of the magnetic energy.
\end{abstract}




\end{frontmatter}


\section{\label{sec:Intro}Introduction}
\label{sec: Intro}
The motion of Lagrangian fluid particles in magnetohydrodynamic (MHD) flows inherently depends on the topology of the magnetic field lines, i.e., smooth curves tangent to the magnetic field, through the Lorentz force. In an ideal medium with no magnetic diffusion, magnetic field lines evolve as regular material curves: they are advected by the fluid velocity as if they were frozen to the MHD flow. Contrarily, in the presence of magnetic diffusion (finite resistivity), strong gradients in the magnetic field typically give rise to experimentally observable current filaments or sheets, i.e., thin regions with high current density \citep{low1988spontaneous, galsgaard2004three, santos2021diffusion}. Induced by the diffusion of the magnetic vector field, these sheets have far-reaching consequences on the transport of fluid particles as well as on the topology of the magnetic field lines \citep{biskamp1986magnetic, carbone1990coherent, cowley1997current}. Understanding the fundamental processes involved in the magnetic diffusion is crucial in describing the material transport in MHD flows. Moreover, spatial structures arising in intermittent turbulence influence the dissipation, heating, transport and acceleration of charged particles both in laboratory and astrophysical plasmas \citep{matthaeus2015intermittency}. \\ \\
Existing coherent structure diagnostics in MHD flows rely on individual snapshots of the velocity and magnetic fields \citep{zhdankin2013statistical, rempel2013coherent}. Such an Eulerian description, however, fails to highlight important transport barriers that are active over a longer time interval. So far, a Lagrangian analysis in MHD flows has purely been limited to computing advective Lagrangian Coherent Structures (LCSs) \citep{falessi2015lagrangian, rempel2017objective, chian2019supergranular, rempel2023lagrangian}. These govern primarily the transport and mixing of Lagrangian fluid particles \citep{Haller2023}. In ideal MHD flows (with zero viscosity $\nu$ and resistivity $\mu$), magnetic field lines evolve as material vectors and hence are indeed tied to advective LCSs. However, in MHD flows with non-zero magnetic diffusion, the magnetic field vectors no longer evolve as regular material vectors, i.e., their tangent vectors do not satisfy the equations of variations
\begin{equation}
\boldsymbol{\dot{\xi}}(t) = \left[\nabla \boldsymbol{v}\left( \boldsymbol{x}(t),t \right)\right]\boldsymbol{\xi}(t), \quad \boldsymbol{x} \in \mathbb{R}^3, \quad t \in \mathbb{R},
\end{equation} where $ \boldsymbol{\xi} $ is a material vector attached to the trajectory $ \boldsymbol{x}\left(t\right) $ generated by the velocity field  $ \boldsymbol{v}(\boldsymbol{x},t) $. Advective LCSs, therefore, are generally insufficient for describing the diffusive transport of the magnetic field vector. Furthermore, LCS detection tools have frequently been employed to visualize invariant manifolds of the instantaneous magnetic field \citep{rempel2013coherent, borgogno2011barriers, rubino2015detection}. Transport barriers obtained in this fashion are, however, not material in unsteady flows. \\ \\
Here we seek active barriers to the diffusive transport of the magnetic field that have an observable impact on the fluid. 
These barriers are, therefore, physical features that are intrinsic to the MHD fluid. As such, they need to be indifferent to the choice of the frame of reference \citep{Lugt1979}, so that two observers that are related to each other via non-relativistic, Euclidian frame changes of the type
\begin{equation}
\boldsymbol{x} = \mathbf{Q}(t)\boldsymbol{y}+\boldsymbol{b}(t), \quad \mathbf{Q}(t) \in SO(3), \quad  \boldsymbol{b}(t) \in \mathbb{R}^3, \quad \mathbf{Q}^T(t)\mathbf{Q}(t) = \mathbf{I}^{3 \times 3}, \label{eq: FrameChange}
\end{equation} identify the exact same material barriers. Here $ \boldsymbol{b}(t) $ is a time-dependent translation and $ \mathbf{Q}(t) $ is a time-dependent rotation matrix. Objectivity is, therefore, a minimal self-consistency requirement for experimentally reproducible coherent structure diagnostics both in MHD and Navier-Stokes flows. \\ \\
The present work builds upon the recent theory developed in \cite{Haller2020} which seeks frame-indifferent (objective) material barriers to the transport of active vectorial quantities in 2D and 3D Navier-Stokes flows. Examples of such active vector fields in fluids include the vorticity and the linear momentum. Specifically, barriers to the diffusive transport of linear momentum, give rise to observable coherent structures in wall-bounded turbulence \citep{aksamit2022objective} and Rayleigh-Bénard flows \citep{aksamit2023interplay}. \\ \\
For MHD flows, the magnetic field qualifies as a dynamically active vector since it contributes to the linear momentum equation through the Lorentz force. In our work, we seek frame-indifferent material barriers to the diffusive transport of the magnetic field. These objective transport barriers are special material surfaces across which the magnetic diffusion vanishes pointwise. \\ \\
The outline of this paper is as follows. In section \ref{sec: Methods}, we first introduce our set-up and notation. We then discuss advective transport barriers and review relevant aspects of LCSs. Subsequently, we derive Eulerian and Lagrangian barriers to the diffusive transport of the magnetic field. In section \ref{sec: Results}, we compute such active magnetic field barriers for 2D and 3D homogenous isotropic turbulence data as streamcurves of a particular vector field. We offer a systematic comparison between magnetic barriers, linear momentum barriers and advective LCSs in MHD flows.

\section{Methods}
\label{sec: Methods}
We consider a 3D electrically conducting fluid with velocity field $ \boldsymbol{v}(\boldsymbol{x}, t) $ and magnetic field $ \mathbf{B}(\boldsymbol{x}, t) $  known at spatial locations $ \boldsymbol{x} \in U \in \mathbb{R}^3 $ in a bounded invariant set $ U $ at times $ t \in [t_0, t_1] $. In the non-relativistic (low-frequency) regime, the forced MHD fluid satisfies the set of visco-resistive MHD equations \citep{Davidson2002, Griffiths2005}
\begin{align}
\frac{D\boldsymbol{v}}{Dt}(\boldsymbol{x}, t) &= -\nabla p(\boldsymbol{x}, t) + \nu\Delta \boldsymbol{v}(\boldsymbol{x},t) + \mathbf{f}_{L}\left( \boldsymbol{x},t \right) + \mathbf{f}_{ext}(\boldsymbol{x}, t), \label{eq: v_dot}\\ 
\frac{D\mathbf{B}}{Dt}(\boldsymbol{x}, t) &= \left[\boldsymbol{\nabla v}(\boldsymbol{x}, t)\right] \mathbf{B}(\boldsymbol{x}, t) + \eta \Delta \mathbf{B}(\boldsymbol{x}, t), \label{eq: B_dot}\\
\nabla \cdot \mathbf{B}(\boldsymbol{x},t) &= 0 \label{eq: Gauss law}, \quad \boldsymbol{x} \in U \subset \mathbb{R}^3, \quad t \in [t_0, t_1] \subset \mathbb{R},
\end{align} where $ p $ is the total pressure field, $ \mathbf{f}_{L} $ is the Lorentz force, $ \mathbf{f}_{ext} $ is an external force, and $ \nu $ and $ \eta $ respectively denote the kinematic viscosity and the magnetic diffusivity of the fluid. The diffusive term originates from Ohm's constitutive material law
\begin{equation}
\mathbf{E}\left(\boldsymbol{x},t \right)+\boldsymbol{v}\left(\boldsymbol{x},t \right) \times \mathbf{B}\left(\boldsymbol{x},t \right) = \eta \boldsymbol{j}\left(\boldsymbol{x},t \right),
\end{equation} where $ \mathbf{E}\left(\boldsymbol{x},t \right) $ is the electric field and 
\begin{equation}
\boldsymbol{j}\left(\boldsymbol{x},t \right) = \nabla \times \mathbf{B}\left(\boldsymbol{x},t \right)
\end{equation}
is the electric current density vector. The Lorentz force
\begin{align}
\mathbf{f}_{L}\left(\boldsymbol{x},t\right) &= \left(\nabla \times \mathbf{B}(\boldsymbol{x},t) \right) \times \mathbf{B}(\boldsymbol{x},t)
\end{align} couples the magnetic field to the equations of motion of the fluid particle. The magnetic field vector appears on the right hand side of the linear momentum equation (\ref{eq: v_dot}) and actively controls the dynamics of the velocity field. \\ \\
Lagrangian particle trajectories generated by the velocity field $ \boldsymbol{v}$ are solutions of the differential equation
\begin{equation}
\boldsymbol{\dot{x}}(t)= \boldsymbol{v}(\boldsymbol{x}, t).
\end{equation} 
We denote the time-$t$ position of a trajectory starting from $ \boldsymbol{x}_0 $ at time $ t_0 $ by
\begin{equation}
\boldsymbol{x}(t;t_0,\boldsymbol{x}_0) := \mathbf{F}_{t_0}^t(\boldsymbol{x}_0),
\end{equation} where $ \mathbf{F}_{t_0}^t $ is the flow map induced by $ \boldsymbol{v}(\boldsymbol{x},t) $. A material surface $ \mathcal{M}(t) \subset U $ is a time dependent codimension-one manifold transported
by the flow map from its initial position $ \mathcal{M}_0 := \mathcal{M}(t_0) $ as 
\begin{equation} 
\mathcal{M}(t) = \mathbf{F}_{t_0}^t\left(\mathcal{M}_0\right). \label{eq: EvolvingMaterialSurface}
\end{equation} The material surface is uniquely determined by its unit normal vector $ \boldsymbol{n}:=\boldsymbol{n}(\boldsymbol{x},t) $ at each point $ \boldsymbol{x} $ at time $ t $.
\subsection{Advective Barriers}
\label{subsec: AdvectiveBarriers}
Advective barriers are passive material surfaces, whose evolution does not directly change the dynamics of the velocity field. Defining material barriers to advective transport is challenging because all material surfaces $ \mathcal{M}(t) $ perfectly inhibit the transport of passive tracers. In contrast, LCSs are distinguished material surfaces that act as centerpieces to the material deformation and thereby maintain coherence over a sustained time interval $ [t_0,t_1] $ (see \cite{Haller2015}). Hyperbolic LCSs are locally the most repelling or attracting codimension-one material surfaces over a finite time interval. Attracting and repelling LCSs mimic unstable and stable invariant manifolds in autonomous dynamical systems, whereas Elliptic LCSs are closed and shear-maximizing material surfaces analogous to KAM-tori \citep{blazevski2014hyperbolic}. Repelling, attracting and elliptic LCSs converge to classic stable, unstable and elliptic (KAM-tori) invariant manifolds if such manifolds exist in the infinite time limit \citep{Haller2015, Haller2023}. \\ \\
The right Cauchy-Green strain tensor associated with the flow map $ \mathbf{F}_{t_0}^{t_1}\left(\boldsymbol{x}_0\right) $ is defined as \citep{truesdell2004non}
\begin{equation}
\mathbf{C}_{t_0}^{t_1}\left(\boldsymbol{x}_0\right) = \left[ \nabla \mathbf{F}_{t_0}^{t_1}\left(\boldsymbol{x}_0\right)\right]^T \nabla \mathbf{F}_{t_0}^{t_1}\left(\boldsymbol{x}_0\right).
\end{equation} This symmetric and positive definite tensor encodes the Lagrangian deformation in the fluid over the finite time interval $ \left[ t_0, t_1 \right] $ at $ \boldsymbol{x}_0 $. 
To visualize hyperbolic LCSs from trajectories generated from a 3D vector field $ \boldsymbol{v}\left( 
\boldsymbol{x}, t\right) $ over a finite time window $  [t_0, t_1] $, we define the finite time Lyapunov exponent
\begin{equation}
\mathrm{FTLE}_{t_0}^{t_1}\left(\boldsymbol{x}_0\right)=\frac{1}{2|t_1-t_0|}\lambda_{max}\left(\mathbf{C}_{t_0}^{t_1}\left(\boldsymbol{x}_0\right)\right), \label{eq: FTLE}
\end{equation} where $ 
\lambda_{max}\left(\mathbf{C}_{t_0}^{t_1}\left(\boldsymbol{x}_0\right)\right) $ is the largest eigenvalue of $ \mathbf{C}_{t_0}^{t_1}\left(\boldsymbol{x}_0\right) $. The $ \mathrm{FTLE}_{t_0}^{t_1} $ field is an objective Lagrangian diagnostic that measures locally the largest material stretching rate in the flow. Ridges of the $ \mathrm{FTLE}_{t_0}^{t_1} $ field obtained from a forward trajectory integration ($ t_1-t_0 > 0 $) mark initial positions of repelling LCSs (generalized stable manifolds) at time $ t_0 $. Similarly, ridges of the backwards $ \mathrm{FTLE}_{t_0}^{t_1} $ field ($ t_1-t_0 < 0 $) denote initial positions of attracting LCSs (generalized unstable manifolds) at time $ t_0 $. \\ \\
In order to detect elliptic LCSs, we employ the Lagrangian-averaged vorticity deviation (LAVD), by \cite{Haller2016} defined over the finite time interval $ \left[ t_0, t_1\right] $ as
\begin{equation}
\mathrm{LAVD}_{t_0}^{t_1}\left( \boldsymbol{x}_0 \right) = \frac{1}{\left| t_1-t_0\right|} \int_{t_0}^{t_1} \left| \boldsymbol{\omega}\left( \mathbf{F}_{t_0}^s\left( \boldsymbol{x}_0\right), s \right)- \hat{\boldsymbol{\omega}}\left( s\right)\right| \ ds, \label{eq: LAVD}
\end{equation} where $ \boldsymbol{\omega} $ denotes the vorticity and the overhat indicates spatial averaging over a fixed flow domain $ U $. The LAVD depends on the choice of the domain over which the spatial averaging is performed. Once $ U $ is fixed, the LAVD is objective, i.e. frame-invariant with respect to frame changes of the type (\ref{eq: FrameChange}). For our computations, the domain is set to be the full computational domain as also done in \cite{aksamit2023interplay}. In 2D, elliptic LCSs at time $ t_0 $ correspond to the outermost convex level sets of the $ \mathrm{LAVD}_{t_0}^{t_1}\left( \boldsymbol{x}_0 \right) $ field surrounding a unique local maximum.  This definition can also be extended to 3D flows, where elliptic LCSs are identified as toroidal iso-surfaces of the $ \mathrm{LAVD}_{t_0}^{t_1}\left( \boldsymbol{x}_0 \right) $ field surrounding a codimension-two ridge \citep{Neamtu-Halic2019}.
\subsection{Active Magnetic Barriers}
\label{subsec: ActiveMagneticBarriers}
The magnetic flux through a control surface $ \mathcal{M}(t) $ is commonly defined as the surface integral of the normal component of the magnetic field $ \mathbf{B} $ over that surface \citep{Davidson2002, Griffiths2005}
\begin{equation}
\mathrm{Flux}_{\mathbf{B}}(\mathcal{M}(t)) = \int_{\mathcal{M}(t)} \mathbf{B}(\boldsymbol{x}(t), t)\cdot \boldsymbol{n}(\boldsymbol{x},t) \ dA \label{eq: FluxB},
\end{equation} where $ \boldsymbol{n}(\boldsymbol{x},t) $ is the normal vector to the surface $ \mathcal{M}(t) $. The definition of magnetic flux given in formula (\ref{eq: FluxB}), however, suffers from multiple limitations for the purposes of defining an objective intrinsic diffusive flux through a co-moving material surface over the time-interval $ [t_0,t_1] $.
First of all, $ Flux_{\boldsymbol{B}} $ measures the net number of magnetic field lines crossing $ \mathcal{M}(t) $, as opposed to the rate at which magnetic field is transported through $ \mathcal{M}(t) $. Secondly, $ Flux_{\boldsymbol{B}} $ is an instantaneous quantity, and therefore fails to quantify the overall transport of the magnetic field across a material surface $ \mathcal{M}(t) $ over the time-interval $ \left[ t_0, t_1\right] $. Finally, the flux (i.e. the transport rate) of a physical quantity through a surface should have the units of that quantity divided by time and multiplied by the surface area. This is not the case for the existing magnetic flux definition from formula (\ref{eq: FluxB}), because $ \mathrm{Flux}_{\mathbf{B}} $ has the units of $ \mathbf{B} $ times the surface area. \\ \\
To resolve these ambiguities, we introduce the diffusive flux of the magnetic field vector through a material surface $ \mathcal{M}(t) $ over the time-interval $ \left[t_0,t_1\right] $ following the approach by \cite{Haller2020}. Specifically, the transport equation for the magnetic field vector (\ref{eq: B_dot}) can be decomposed into a diffusive and non-diffusive component
\begin{equation}
\frac{D}{Dt}\mathbf{B}(\boldsymbol{x},t) = \mathbf{b}_{non-dif}(\boldsymbol{x},t)+\mathbf{b}_{dif}(\boldsymbol{x},t),
\end{equation} with $ \mathbf{b}_{dif}(\boldsymbol{x},t) = \eta \Delta \mathbf{B}(\boldsymbol{x},t) $ and $ \mathbf{b}_{non-dif}(\boldsymbol{x},t) = \left[ \mathbf{\nabla} \boldsymbol{v}(\boldsymbol{x},t) \right] \mathbf{B}(\boldsymbol{x},t) $. Magnetic field lines are transported either via magnetic diffusion ($ \mathbf{b}_{dif} $) or advection ($ \mathbf{b}_{non-dif} $). In the absence of magnetic diffusion, the magnetic field lines are advected by the velocity field as regular material lines. \\ \\
We now introduce the instantaneous (Eulerian) diffusive flux of the magnetic field through a material surface $ \mathcal{M}(t) $ as
\begin{align}
\Phi(\mathcal{M}(t)) &= \int_{\mathcal{M}(t)} \left[ \frac{D}{Dt}\mathbf{B}(\boldsymbol{x},t) \cdot \boldsymbol{n}(\boldsymbol{x},t)\right]_{dif}dA
\label{eq: LagrangianDiffusiveFlux}\\
&= \int_{\mathcal{M}(t)} \mathbf{b}_{dif}(\boldsymbol{x},t) \cdot \boldsymbol{n}(\boldsymbol{x},t) \ dA \notag
\end{align}
Physically, $ \Phi $ quantifies the time-normalized transport of the magnetic field across a material surface $ \mathcal{M}(t) $ due to diffusion. As expected, the functional $ \Phi $ has no explicit dependence on the velocity field. Note that fluid trajectories do not even need to physically cross the surface $ \mathcal{M}(t)$ to induce a diffusive magnetic flux. Compared to the classic magnetic flux $ Flux_{\boldsymbol{B}} $ (see formula (\ref{eq: FluxB})), the newly introduced diffusive magnetic flux $ \Phi $ has the physical units expected for the flux of the magnetic field vector: It is given by the units of the magnetic field multiplied by area and divided by time. \\ \\
Eventhough $ Flux_{\boldsymbol{B}} $ fails to satisfy the physical requirements of a flux, we remain consistent with the cited references in the MHD literature and denote $ Flux_{\boldsymbol{B}} $ as the magnetic flux and $ \Phi $ as the diffusive magnetic flux. Alternatively, we can obtain $ \Phi $ by taking the material derivative of $ Flux_{\boldsymbol{B}} $ \citep{stern1966motion, eyink2006breakdown}:
\begin{align}
\frac{D}{Dt} Flux_{\boldsymbol{B}}\left( \mathcal{M}(t) \right) &=
\int_{\mathcal{M}(t)}\left(\dfrac{\partial}{\partial t}\boldsymbol{B}\left( \boldsymbol{x},t \right) -\nabla \times \left( \boldsymbol{v}\left( \boldsymbol{x},t \right) \times \boldsymbol{B}\left( \boldsymbol{x},t \right) \right) \right) \boldsymbol{n}\left( \boldsymbol{x},t \right) dA \\
&= \int_{\mathcal{M}(t)}\left(\dfrac{\partial}{\partial t}\boldsymbol{B}\left( \boldsymbol{x},t \right) + \left[\nabla \boldsymbol{B}\left( \boldsymbol{x},t \right) \right] \boldsymbol{v}\left( \boldsymbol{x},t \right) - \left[\nabla \boldsymbol{v}\left( \boldsymbol{x},t \right) \right] \boldsymbol{B}\left( \boldsymbol{x},t \right) \right) \boldsymbol{n}\left( \boldsymbol{x},t \right) dA \\
&= \int_{\mathcal{M}(t)}\left(\dfrac{D}{Dt}\boldsymbol{B}\left( \boldsymbol{x},t \right) - \left[\nabla \boldsymbol{v}\left( \boldsymbol{x},t \right) \right] \boldsymbol{B}\left( \boldsymbol{x},t \right) \right) \boldsymbol{n}\left( \boldsymbol{x},t \right) dA \\
&= \eta \int_{\mathcal{M}(t)} \Delta \boldsymbol{B}\left( \boldsymbol{x},t \right) \boldsymbol{n}\left( \boldsymbol{x},t \right) dA \\
&= \Phi \left(\mathcal{M}(t)\right). \label{eq: AlfvenDiffusive}
\end{align}
From formula (\ref{eq: AlfvenDiffusive}), we see that the rate-of-change of $ Flux_{\boldsymbol{B}} $ coincides with the diffusive magnetic flux $ \Phi $. This implies that for a given material surface $ \mathcal{M}(t) $, any change in $ Flux_{\mathbf{B}} $ can only occur due to diffusive process. According to Alfvén's Theorem \cite{alfven1943existence}, in ideal MHD fluid ($\eta = 0 $), the magnetic flux $ Flux_{\boldsymbol{B}} $ is conserved along any arbitrary material surface $ \mathcal{M}(t) $:
\begin{equation}
\dfrac{D}{Dt} Flux_{\boldsymbol{B}}(\mathcal{M}(t)) = 0 \label{eq: Alfven Theorem}.
\end{equation} 
However, in non-ideal MHD flows ($\eta > 0 $), $ Flux_{\boldsymbol{B}} $ is conserved only for a specific set of $ \mathcal{M}(t) $, where the diffusive magnetic flux $ \Phi $ vanishes. \\ \\
In order to obtain the diffusive Lagrangian magnetic flux, we integrate the Eulerian flux $ \Phi(\mathcal{M}(t)) $ along trajectories defining the evolving material surface $ \mathcal{M}(t) $, which yields
\begin{align}
\Psi_{t_0}^{t_1}(\mathcal{M}(t)) &= \frac{1}{t_1-t_0}\int_{t_0}^{t_1}\Phi(\mathcal{M}(t)) \ dt \label{eq: Lagrangian_flux_integration_Eulerian_flux} \\
&= \frac{1}{t_1-t_0}\int_{t_0}^{t_1}\int_{\mathcal{M}(t)}\boldsymbol{b}_{dif}(\boldsymbol{x},t) \cdot \boldsymbol{n}(\boldsymbol{x},t) \ dAdt \notag .
\end{align}
Under non-relativistic Euclidian frame changes of the form (\ref{eq: FrameChange}), the magnetic field vector is a frame-indifferent vector field, because it transforms as an objective vector \citep{muller2023electrodynamics}
\begin{equation}
\mathbf{\tilde{B}}\left(\boldsymbol{y},t\right) = \mathbf{Q}^T(t)\mathbf{B}\left(\boldsymbol{x},t \right) \label{eq: Transformed_magnetic}.
\end{equation} Since the rotation matrix $ \mathbf{Q}(t) $ has no spatial dependence, it remains unaffected by spatial differentiation and the Laplacian of $ \mathbf{B} $ is also guaranteed to transform properly $ \Delta \mathbf{\tilde{B}}\left(\boldsymbol{y},t\right) = \mathbf{Q}^T(t) \Delta \mathbf{B}\left(\boldsymbol{x},t \right) $. Under an observer change of the form (\ref{eq: FrameChange}), the transformation formula \[ \boldsymbol{\tilde{n}}(\boldsymbol{y},t) = \mathbf{Q}^T(t) \boldsymbol{n}(\boldsymbol{x},t) \] for the unit normal vector $ \boldsymbol{n} $ implies that both the Eulerian and the Lagrangian diffusive transport of the magnetic field vector are objective because
\begin{align}
\tilde{\Phi}(\mathcal{\tilde{M}}(t)) &= \int_{\mathcal{\tilde{M}}(t)}\boldsymbol{\tilde{b}}_{dif}(\boldsymbol{y},t)\cdot \boldsymbol{\tilde{n}}(\boldsymbol{y},t) \ d\tilde{A} \notag \\
&= \int_{\mathcal{M}(t)} \left(\mathbf{Q}^T(t)\boldsymbol{b}_{dif}(\boldsymbol{x},t)\right)\cdot\left( \mathbf{Q}^T(t)\boldsymbol{n}(\boldsymbol{x},t)\right) \ dA \notag \\
&= \int_{\mathcal{M}(t)} \boldsymbol{b}_{dif}\left( \boldsymbol{x},t\right)\cdot \boldsymbol{n}(\boldsymbol{x},t) \ dA \notag \\
&= \Phi(\mathcal{M}(t)),
\end{align} and similarly
\begin{align}
\tilde{\Psi}_{t_0}^{t_1}(\mathcal{\tilde{M}}(t))
= \Psi_{t_0}^{t_1}(\mathcal{M}(t)).
\end{align}
Thanks to its inherent frame-indifference, the Eulerian and Lagrangian active magnetic barriers can be thought of as intrinsic physical properties of the surface and flow. Indeed, they are specifically tied to the fluid and do not depend on the reference frame of the observer. \\ \\
Using the surface-element deformation formula \citep{gurtin2010mechanics}
\begin{equation}
\boldsymbol{n}(\boldsymbol{x},t)dA = \mathrm{det}\left( \mathbf{\nabla F}_{t_0}^t(\boldsymbol{x}_0)\right)\left[ \mathbf{\nabla F}_{t_0}^t(\boldsymbol{x}_0) \right]^{-T}\boldsymbol{n}(\boldsymbol{x}_0) \ dA_0,
\end{equation} we can parametrize the functional $ \Psi_{t_0}^t $ over its initial material surface $ \mathcal{M}_0 $ as
\begin{align}
\Psi_{t_0}^{t_1}(\mathcal{M}_0) &= \frac{1}{t_1-t_0}\int_{t_0}^{t_1}\int_{\mathcal{M}_0}\boldsymbol{b}_{dif}(\mathbf{F}_{t_0}^t\left(\boldsymbol{x}_0 \right),t) \cdot \mathrm{det}\left( \mathbf{\nabla F}_{t_0}^t(\boldsymbol{x}_0)\right)\left[ \mathbf{\nabla F}_{t_0}^t(\boldsymbol{x}_0) \right]^{-T} \boldsymbol{n}(\boldsymbol{x}_0) \ dA_0dt \notag \\
&= \int_{\mathcal{M}_0} \frac{1}{t_1-t_0}\int_{t_0}^{t_1} \mathrm{det}\left( \mathbf{\nabla F}_{t_0}^t(\boldsymbol{x}_0)\right)\left[ \mathbf{\nabla F}_{t_0}^t(\boldsymbol{x}_0) \right]^{-1}\boldsymbol{b}_{dif}(\mathbf{F}_{t_0}^t\left(\boldsymbol{x}_0 \right),t) dt \cdot \boldsymbol{n}(\boldsymbol{x}_0) \ dA_0 \notag \\
&= \int_{\mathcal{M}_0} \boldsymbol{\overline{b}}_{t_0,dif}^{t_1}(\boldsymbol{x}_0) \cdot \boldsymbol{n}(\boldsymbol{x}_0) \ dA_0,  \label{eq: LagrangianFluxM0}
\end{align} with 
\begin{equation}
\boldsymbol{\overline{b}}_{t_0,dif}^{t_1}(\boldsymbol{x}_0) = \frac{1}{t_1-t_0} \int_{t_0}^{t_1} \mathrm{det}\left( \mathbf{\nabla F}_{t_0}^t(\boldsymbol{x}_0)\right)\left[ \mathbf{\nabla F}_{t_0}^t(\boldsymbol{x}_0) \right]^{-1}\boldsymbol{b}_{dif}(\mathbf{F}_{t_0}^t\left(\boldsymbol{x}_0 \right),t) \ dt. \label{eq: h_diff_Lagrangian_init}
\end{equation}
Later positions of $ \mathcal{M}_0 $ can be obtained through material advection (see relationship (\ref{eq: EvolvingMaterialSurface})).
To keep our notation simple, we denote the temporal average of a Lagrangian vector field $ \boldsymbol{w}(\boldsymbol{x}_0,t) $ as
\begin{equation}
\boldsymbol{\overline{w}}(\boldsymbol{x}_0) = \frac{1}{t_1-t_0}\int_{t_0}^{t_1} \boldsymbol{w}(\boldsymbol{x}_0,t) \ dt. \label{eq: Average}
\end{equation} We also introduce $ \left(\mathbf{F}_{t_0}^{t}\right)^{*}\boldsymbol{u}(\boldsymbol{x}_0) $ as the pullback operator \citep{do1998differential} of an Eulerian vector field $ \boldsymbol{u}(\boldsymbol{x},t) $ under the flow map $ \mathbf{F}_{t_0}^t $ to the initial configuration at $ t_0 $
\begin{equation}
\left(\mathbf{F}_{t_0}^{t}\right)^{*} \left[\boldsymbol{u}(\boldsymbol{x}_0)\right] = \left[ \nabla \mathbf{F}_{t_0}^t\right]^{-1}\boldsymbol{u}(\mathbf{F}_{t_0}^t(\boldsymbol{x}_0),t) \label{eq: Pullback}.
\end{equation}
Using the notation (\ref{eq: Average})-(\ref{eq: Pullback}), we obtain the simplified expression
\begin{equation}
\boldsymbol{\overline{b}}_{t_0,dif}^{t_1}(\boldsymbol{x}_0) = \overline{\mathrm{det}\left( \nabla \mathbf{F}_{t_0}^{t}(\boldsymbol{x}_0)\right) \left(\mathbf{F}_{t_0}^{t}\right)^{*} \left[\boldsymbol{b}_{dif}\left(\boldsymbol{x}_0 \right)\right]} \label{eq: h_diff_Lagrangian},
\end{equation} and we rewrite formula (\ref{eq: LagrangianFluxM0}) as
\begin{equation}
\Psi_{t_0}^{t_1}\left(\mathcal{M}_0\right) = \int_{\mathcal{M}_0} \boldsymbol{\overline{b}}_{t_0,dif}^{t_1}(\boldsymbol{x}_0) \cdot \boldsymbol{n}(\boldsymbol{x}_0) \ dA_0 \label{eq: LagrangianFluxM0_2}.
\end{equation}
We seek active Lagrangian barriers to the magnetic vector field as material surfaces $ \mathcal{M}_0 $ along which the integrand in the transport functional $ \Psi_{t_0}^{t_1} $ vanishes pointwise. This occurs if $ \mathcal{M}_0 $ is everywhere tangent to $ \boldsymbol{\overline{b}}_{t_0,dif}^{t_1}(\boldsymbol{x}_0) $. These material surfaces necessarily coincide with streamsurfaces (i.e., codimension-one invariant manifolds) of the autonomous vector field $ \boldsymbol{\overline{b}}_{t_0,dif}^{t_1}(\boldsymbol{x}_0) $.
We parametrize the streamlines $ \boldsymbol{x}_0 $ of $ \boldsymbol{\overline{b}}_{t_0,dif}^{t_1}\left(\boldsymbol{x}_0\right) $ with $ s \in \mathbb{R} $, i.e., they satisfy the differential equation
\begin{equation}
\boldsymbol{x}_0'(s) = \boldsymbol{\overline{b}}_{t_0,dif}^{t_1}\left(\boldsymbol{x}_0(s)\right) \label{eq: Lagrangian_barrier_first},
\end{equation} where prime denotes differentiation with respect to $ s $. By taking the limit $ t_1 \rightarrow t_0 = t $ in formula (\ref{eq: Lagrangian_flux_integration_Eulerian_flux}), we obtain the diffusive Eulerian magnetic flux
\begin{equation}
\lim\limits_{t_1 \rightarrow t_0 = t} \Psi_{t_0}^{t_1}(\mathcal{M}_0) := \Phi(\mathcal{M}(t)) = \int_{\mathcal{M}(t)} \boldsymbol{b}_{dif}\left( \boldsymbol{x},t\right)\cdot \boldsymbol{n}(\boldsymbol{x},t) \ dA,
\end{equation} where $ \lim\limits_{t_0\rightarrow t_1 =t} \boldsymbol{\overline{b}}_{t_0,dif}^{t_1}(\boldsymbol{x}_0):=\boldsymbol{b}_{dif}\left(\boldsymbol{x},t \right) $ and $ \lim\limits_{t_0\rightarrow t_1 = t} \nabla \boldsymbol{F}_{t_0}^t\left(\boldsymbol{x}_0\right):= \boldsymbol{I} $. Therefore, material surfaces $ \mathcal{M}(t) $ minimizing $ \Phi $ coincide with streamsurfaces $ \boldsymbol{x} $ of $ \boldsymbol{b}_{dif}(\boldsymbol{x},t) $
\begin{equation}
\boldsymbol{x}'(s) = \boldsymbol{b}_{dif}(\boldsymbol{x}(s),t) \label{eq: Eulerian_barrier_first}.
\end{equation}
Using the formulas (\ref{eq: Lagrangian_barrier_first})-(\ref{eq: Eulerian_barrier_first}), leads us to formulating the following definition.
\begin{definition}
\label{def: Definition1}
For electrically conducting fluid flows, exact Eulerian and Lagrangian barriers to the diffusive (resistive) transport of the magnetic field are invariant manifolds of
\begin{align}
\boldsymbol{x}'(s) &= \eta \Delta \mathbf{B}(\boldsymbol{x}(s),t), \label{eq: Theorem1_Eulerian} \\
\boldsymbol{x}_0'(s) &= \eta \overline{\mathrm{det}\left( \nabla \mathbf{F}_{t_0}^{t}(\boldsymbol{x}_0)\right) \left(\mathbf{F}_{t_0}^{t}\right)^{*} \left[\Delta \mathbf{B}(\boldsymbol{x}_0(s)) \right]}. \label{eq: Theorem1_Lagrangian}
\end{align}
\end{definition}
The barrier fields (\ref{eq: Theorem1_Eulerian})-(\ref{eq: Theorem1_Lagrangian}) define a 3D, autonomous (or steady) dynamical system that remain valid also for spatially and temporally dependent magnetic diffusivity $ \left(\eta := \eta\left(\boldsymbol{x}, t\right)\right)  $. We can parametrize the trajectories generated by the barrier field (\ref{eq: Theorem1_Eulerian}) with respect to the rescaled barrier time $ \tau $
\begin{equation} 
\tau(s) = \int_0^s\eta(x(\tilde{s}))d\tilde{s}.
\end{equation}
An analogous statement also holds for the trajectories satisfying the active Lagrangian barrier field (\ref{eq: Theorem1_Lagrangian}). This implies that the topology of the barrier fields (\ref{eq: Theorem1_Eulerian})-(\ref{eq: Theorem1_Lagrangian}) is not affected by temporally and spatially dependent magnetic diffusivity. \\ \\
Note that even the Lagrangian barrier field (\ref{eq: Theorem1_Lagrangian}) is a steady vector field once we fix the initial $ \left( t_0 \right) $ and final $ \left(t_1\right) $ time. All the relevant information about the time evolution of $ \boldsymbol{v}(\boldsymbol{x}, t) $ and $ \mathbf{B}(\boldsymbol{x}, t) $ over the time interval $ [t_0, t_1] $ is encoded in the pullback and the temporal averaging operations. The instantaneous version (\ref{eq: Theorem1_Eulerian}) only contains the physical time $ t $ as a single parameter. The Eulerian barrier field is always a divergence free vector field because $ \mathbf{B} $ is divergence free due to Gauss's law for magnetism. This holds even for compressible fluids. In contrast, in compressible flows, the Lagrangian barrier field (\ref{eq: Theorem1_Lagrangian}) is generally not divergence free. \\ \\
By construction, all of the trajectories of the barrier fields (\ref{eq: Theorem1_Eulerian})-(\ref{eq: Theorem1_Lagrangian}) are exact barriers to the Eulerian (or Lagrangian) diffusive magnetic flux. Certain transport barriers, however, stand out because of their unique topology (e.g. they are closed and convex), while others because of their strength (e.g. they inhibit large fluxes). To obtain a direct measure of the local strength of an active barrier, we introduce the geometric flux density
\begin{equation}
g\left( \boldsymbol{x}_0; \boldsymbol{f}, \boldsymbol{n} \right) = \left| \boldsymbol{f}\left(\boldsymbol{x}_0\right) \cdot \boldsymbol{n}\left(\boldsymbol{x}_0\right) \right|,
\end{equation} as defined by \cite{mackay1994transport}. Here, $ \boldsymbol{f} $ is a general active barrier field. For example, when treating instantaneous active magnetic barriers, we set 
\begin{equation} 
\boldsymbol{f} = \eta \Delta \mathbf{B},
\end{equation} whereas when computing Lagrangian active magnetic barriers, we set  
\begin{equation}
\boldsymbol{f} = \eta \overline{\mathrm{det}\left( \nabla \mathbf{F}_{t_0}^{t}(\boldsymbol{x}_0)\right) \left(\mathbf{F}_{t_0}^{t}\right)^{*} \left[\Delta \mathbf{B}(\boldsymbol{x}_0(s)) \right]}.
\end{equation} In analogy with the Diffusion Barrier Strength ($ \mathrm{DBS} $) defined in \cite{haller2018material, haller2020barriers}, the local strength of an active barrier field $ \boldsymbol{f} $ is given by the Active Barrier Strength
\begin{equation}
\mathrm{ABS}\left( \boldsymbol{x}_0;\boldsymbol{f}\right) = \left| \boldsymbol{f}\left(\boldsymbol{x}_0\right) \right|. \label{eq: ABS}
\end{equation} This follows from the fact that the geometric flux density $ g\left( \boldsymbol{x}_0; \boldsymbol{f}, \boldsymbol{n} \right) $ will change the most under a small change in the relative position of the vectors $ \boldsymbol{f}\left(\boldsymbol{x}_0\right) $ and $ \boldsymbol{n}\left(\boldsymbol{x}_0\right) $ at locations where $ \left| \boldsymbol{f}\left(\boldsymbol{x}_0\right) \right| $ is the largest. As a result, the $ \mathrm{ABS} $ provides an objective and robust scalar diagnostic field that highlights the most influential active transport barriers. Analogous to ridges of the $ \mathrm{DBS}$ field that highlight the most influential diffusive transport minimizers in the flow, we can detect exceptionally strong active barriers as codimension-one ridges of the $ \mathrm{ABS} $ field \citep{haller2018material, haller2020barriers}.  \\ \\
Note, however, that ridges of the $ \mathrm{ABS} $ field only serve as approximate barriers to the diffusive transport of the magnetic field. In 2D, we can exactly compute the most influential active barriers as streamlines of $ \boldsymbol{f} $ passing through local maxima of the $ \mathrm{ABS} $ field. Since we expect the most influential barriers to be characterized by high $ \mathrm{ABS} $ values, we launch trajectories from local maxima of the $ \mathrm{ABS} $ and stop the trajectory integration once the $ \mathrm{ABS} $ falls below a predefined threshold $ \varepsilon_{\mathrm{ABS}} $. The identified barriers are robust because local maxima and ridges of a scalar field are topologically robust features \citep{karrasch2013finite}, i.e., they persist with respect to small volume-preserving perturbations to the underlying field. Additionally, to filter out small scale barriers linked to noise, we discard trajectory segments shorter than $ \ell_{min} $. For the same reason, we only retain trajectory segments whose minimal distance to neafrby barriers exceeds $ \mathrm{d}_{min} $. Both $ \ell_{min} $ and $ \mathrm{d}_{min} $ typically depend on the turbulent length scale of the MHD flow. The exact computational details are provided in the Algorithm \ref{alg: InfluentialActiveBarrier2D}.
\begin{algorithm}
  \caption{Strongest active barriers from barrier field $ \boldsymbol{f} $ in 2D. \label{alg: InfluentialActiveBarrier2D}}
  \begin{flushleft}
  \textbf{Input: } 2D active barrier field $ \boldsymbol{f}\left( \boldsymbol{x} \right) $ over a regular meshgrid $ \boldsymbol{x} \in U $. \\
  \textbf{Output: } Strongest active barriers blocking the transport of $ \boldsymbol{f}\left( \boldsymbol{x} \right) $.
  \end{flushleft}
  \begin{algorithmic}[1]
  \State Compute Active Barrier Strength \begin{align} \mathrm{ABS}\left( \boldsymbol{x};\boldsymbol{f} \right) &= \left| \boldsymbol{f}\left( \boldsymbol{x}\right) \right|.
  \end{align}
    \State Compute the set $ \mathcal{S}_{loc,max} $ of local maxima of $ \mathrm{ABS}\left( \boldsymbol{x};\boldsymbol{f} \right)$ and sort them in descending order.
    \State Compute the strongest active barriers as trajectories of $ \boldsymbol{f} $ passing through local maxima in the set $ \mathcal{S}_{loc,max} $.
    \State Retain trajectory segments satisfying the following conditions:
    \begin{algsubstates}
        \State The arclength exceeds $ \ell_{min} $.
        \State The pointwise $ \mathrm{ABS} $ is larger than $ \varepsilon_{\mathrm{ABS}} $.
        \State The minimal distance to the nearest active barrier is larger then $ \mathrm{d}_{min} $.
      \end{algsubstates}
  \end{algorithmic}
\end{algorithm} \\ \\
In 2D flows, the active magnetic barriers identified by Algorithm \ref{alg: InfluentialActiveBarrier2D} are obtained as streamlines of the appropriate barrier field $ \boldsymbol{f} $ and are, therefore, exact barriers according to Definition \ref{def: Definition1}. Contrarily, in 3D flows, 2D invariant manifolds of $ \boldsymbol{f} $ can only be determined approximately. To obtain active transport magnetic barriers, we first evaluate the $ \mathrm{ABS} $ field over a cross-section of the selected domain. Each individual ridge of the $ \mathrm{ABS} $ field corresponds to a smooth curve that forms a set of initial conditions for which we compute streamlines of $ \boldsymbol{f} $. Analogously to the 2D case, we only retain trajectory segments whose pointwise $ \mathrm{ABS} $ is greater than $ \varepsilon_{\mathrm{ABS}} $. For each ridge, we obtain a distinguished active transport barrier by fitting a surface through the set of streamlines. These barriers act as 2D surfaces that locally divide the domain into two regions that exchange minimal amounts of $ \boldsymbol{f} $.
\begin{algorithm}
  \caption{Strongest active barriers from barrier field $ \boldsymbol{f} $ in 3D. \label{alg: InfluentialActiveBarrier3D}}
  \begin{flushleft}
  \textbf{Input: } 3D active barrier field $ \boldsymbol{f}\left(\boldsymbol{x} \right) $ over a regular meshgrid $ \boldsymbol{x} \in U $. \\
  \textbf{Output: } Strongest active barriers blocking the transport of $ \boldsymbol{f}\left( \boldsymbol{x} \right) $.
  \end{flushleft}
  \begin{algorithmic}[1]
  \State Compute Active Barrier Strength \begin{align} \mathrm{ABS}\left( \boldsymbol{x};\boldsymbol{f} \right) &= \left| \boldsymbol{f}\left( \boldsymbol{x}\right) \right|.
  \end{align} over a 2D cross-section of the 3D domain.
    \State Compute ridges of $ \mathrm{ABS}\left( \boldsymbol{x};\boldsymbol{f} \right)$ using Lindeberg's ridge extraction algorithm \citep{lindeberg1998edge}. Each ridge corresponds to a smooth curve.
    \State Compute streamlines of $ \boldsymbol{f} $ for each smooth curve (=ridge). Only retain trajectory segments satisfying the following conditions:    \begin{algsubstates}
        \State The arclength does not exceed $ \ell_{max} $.
        \State The pointwise $ \mathrm{ABS} $ is larger than $ \varepsilon_{\mathrm{ABS}} $.
      \end{algsubstates}
    \State The streamlines launched from each ridge form a point cloud. Fit a polynomial surface of degree $ \mathrm{d} $ through the point cloud using regression. The obtained surface acts as an approximate barrier blocking the transport of $ \boldsymbol{f} $.
  \end{algorithmic}
\end{algorithm}
\subsubsection{Two-Dimensional Incompressible MHD fluids}
\label{subsec: IncompressibleMHD}
In volume-preserving (incompressible) MHD fluids we have
\begin{equation}
\mathrm{det}\left( \mathbf{\nabla F}_{t_0}^t(\boldsymbol{x}_0)\right) = 1
\end{equation} We can therefore rewrite the Lagrangian barrier equation (\ref{eq: Theorem1_Lagrangian}) as
\begin{equation}
\boldsymbol{x}_0'(s) = \eta \overline{\left(\mathbf{F}_{t_0}^{t}\right)^{*} \left[\Delta \mathbf{B}(\boldsymbol{x}_0(s)) \right]} \label{eq: LagrangianBarrierIncompressible}.
\end{equation}
In 2D incompressible MHD flows, we can derive analytical expressions for the barrier equations from Definition \ref{def: Definition1}. The visco-resistive MHD equations (\ref{eq: v_dot}-\ref{eq: B_dot}) then reduce to a pair of advection-diffusion equations
\begin{align}
\frac{\partial}{\partial t} \omega \left(\boldsymbol{x}, t \right)+\boldsymbol{v}\left(\boldsymbol{x}, t \right) \nabla \omega \left(\boldsymbol{x}, t \right) &= \nu \Delta \omega \left(\boldsymbol{x}, t \right)+\mathbf{B}\left(\boldsymbol{x}, t \right)\nabla j \left(\boldsymbol{x}, t \right), \label{eq: AdvectionDiffusionVorticity}\\
\frac{\partial}{\partial t} a\left(\boldsymbol{x}, t \right)+\boldsymbol{v}\left(\boldsymbol{x}, t \right) \nabla a\left(\boldsymbol{x}, t \right) &= \eta \Delta a\left(\boldsymbol{x}, t \right), \label{eq: AdvectionDiffusionMagnetic}
\end{align} where $ a(\boldsymbol{x},t) $ is the magnetic scalar potential. The vorticity and the electric current density are both scalars and given by \[ \omega(\boldsymbol{x},t) = -\Delta \psi\left(\boldsymbol{x},t\right) \] and \[ j(\boldsymbol{x},t) = -\Delta a\left(\boldsymbol{x},t\right). \] The streamfunction $ \psi \left(\boldsymbol{x},t\right) $ and the scalar magnetic potential $ a\left(\boldsymbol{x},t\right) $ then act as time-dependent Hamiltonians to the velocity and magnetic field
\begin{align}
\boldsymbol{v}\left( \boldsymbol{x},t \right) &= -\mathbf{J} \nabla \psi\left(\boldsymbol{x},t\right), \\
\mathbf{B}\left(\boldsymbol{x}, t \right)  &= -\mathbf{J} \nabla a\left(\boldsymbol{x}, t \right), \label{eq: MagneticField_Potential}
\end{align} with $ \mathbf{J} = \begin{pmatrix}
0 && -1 \\ 1 && 0
\end{pmatrix} $ and the incompressibility of the magnetic field implies
\begin{align}
\Delta \mathbf{B}\left(\boldsymbol{x}, t \right) = \mathbf{J} \mathbf{\nabla} j_z\left(\boldsymbol{x}, t \right) \label{eq: jz}.
\end{align}
With this notation, we obtain the following results on active magnetic barriers in incompressible 2D MHD flows.
\begin{theorem}
\label{th: Theorem2}
For incompressible, electrically conducting 2D fluid flows, exact Eulerian and Lagrangian barriers to the diffusive (resistive) transport of magnetic field are  invariant manifolds of
\begin{align}
\boldsymbol{x}'(s) &= \eta \mathbf{J} \mathbf{\nabla} j_z(\boldsymbol{x}(s),t), \label{eq: Theorem2_Eulerian} \\
\boldsymbol{x}_0'(s) &= \eta \mathbf{J} \mathbf{\nabla}_0 \overline{j_z\left( \boldsymbol{x}_0(s)\right)} \label{eq: Theorem2_Lagrangian},
\end{align} where $ \overline{j_z\left( \boldsymbol{x}_0 \right)} $ denotes the temporal average over the time-interval $ \left[t_0, t_1 \right] $ of the electric current density along a fluid trajectory $ \boldsymbol{x}\left(t;t_0,\boldsymbol{x}_0\right):= \mathbf{F}_{t_0}^{t}\left( \boldsymbol{x}_0 \right) $.
\end{theorem}
The Eulerian barrier equation (\ref{eq: Theorem2_Eulerian}) directly follows from formula (\ref{eq: jz}). The derivation of the Lagrangian active magnetic barrier equation (\ref{eq: Theorem2_Lagrangian}) is outlined in Appendix \ref{app: Theorem 2}. The scalar functions $ \overline{j_z\left( \boldsymbol{x}_0\right)} $ and $ j_z\left(\boldsymbol{x}_0,t\right) $ act as Hamiltonians to the Lagrangian and Eulerian active magnetic barrier equations (\ref{eq: Theorem2_Eulerian}-\ref{eq: Theorem2_Lagrangian}).
\subsection{Active Linear Momentum Barriers}
\label{subsec: ActiveLinearMomentumBarriers}
Active barriers to the diffusive transport of linear momentum $ (\rho \boldsymbol{v}) $ in Navier-Stokes flows arise due to viscous/diffusive forces in the flow \citep{Haller2020}. Here $ \rho $ is the constant density of the fluid and is universally set to $ 1 $. We can obtain momentum barriers for MHD flows by following the same principles. Specifically, as in our treatment of the active magnetic barriers, we decompose the right hand side of the linear momentum equation in MHD flows (see \ref{eq: v_dot}) into diffusive (viscous) and non-diffusive (non-viscous) components. Here, the only diffusive force in the MHD momentum equation is given by $ \nu \Delta \boldsymbol{v} $ and we obtain for 3D MHD flows the exact same momentum barrier fields as for 3D Navier-Stokes flows:
\begin{align}
\boldsymbol{x}'(s) &= \nu \Delta \boldsymbol{v}(\boldsymbol{x}(s),t), \label{eq: 3D_Eulerian_momentum} \\
\boldsymbol{x}_0'(s) &= \nu \overline{\mathrm{det}\left( \nabla \mathbf{F}_{t_0}^{t}(\boldsymbol{x}_0)\right) \left(\mathbf{F}_{t_0}^{t}\right)^{*} \left[\Delta \boldsymbol{v}(\boldsymbol{x}_0(s)) \right]} \label{eq: 3D_Lagrangian_momentum}.
\end{align}
In 2D MHD flows the linear momentum barrier fields are Hamiltonian and simplify to
\begin{align}
\boldsymbol{x}'(s) &= \nu \mathbf{J} \mathbf{\nabla} \omega(\boldsymbol{x}(s),t), \label{eq: 2D_Eulerian_momentum} \\
\boldsymbol{x}_0'(s) &= \nu \mathbf{J} \mathbf{\nabla}_0 \overline{\omega\left( \boldsymbol{x}_0(s)\right)} \label{eq: 2D_Lagrangian_momentum},
\end{align} where $ \overline{\omega\left( \boldsymbol{x}_0\right)} $ denotes the temporal average over the time-interval $ \left[ t_0, t_1 \right] $ of the vorticity along a fluid trajectory $ \boldsymbol{x}\left(t;t_0,\boldsymbol{x}_0\right):= \mathbf{F}_{t_0}^{t}\left( \boldsymbol{x}_0 \right) $. Note that for the momentum barriers, the vorticity $ \omega $ plays the same role as the electric current density $ j_z $ in the active magnetic barrier equations (\ref{eq: Theorem2_Eulerian})-(\ref{eq: Theorem2_Lagrangian}). For a detailed derivation of the 2D and 3D linear momentum barriers we refer to the original work by \cite{Haller2020}.
\section{Results}
\label{sec: Results}
We now illustrate the numerical implementation of our results on high resolution 2D and 3D MHD turbulence simulations. The codes and the snapshots of the datasets are available in the GitHub repository of
\href{https://github.com/EncinasBartos}{Encinas Bartos}.
\subsection{Two-dimensional MHD Turbulence}
\begin{figure*}
\centering{\includegraphics[width=1\textwidth]{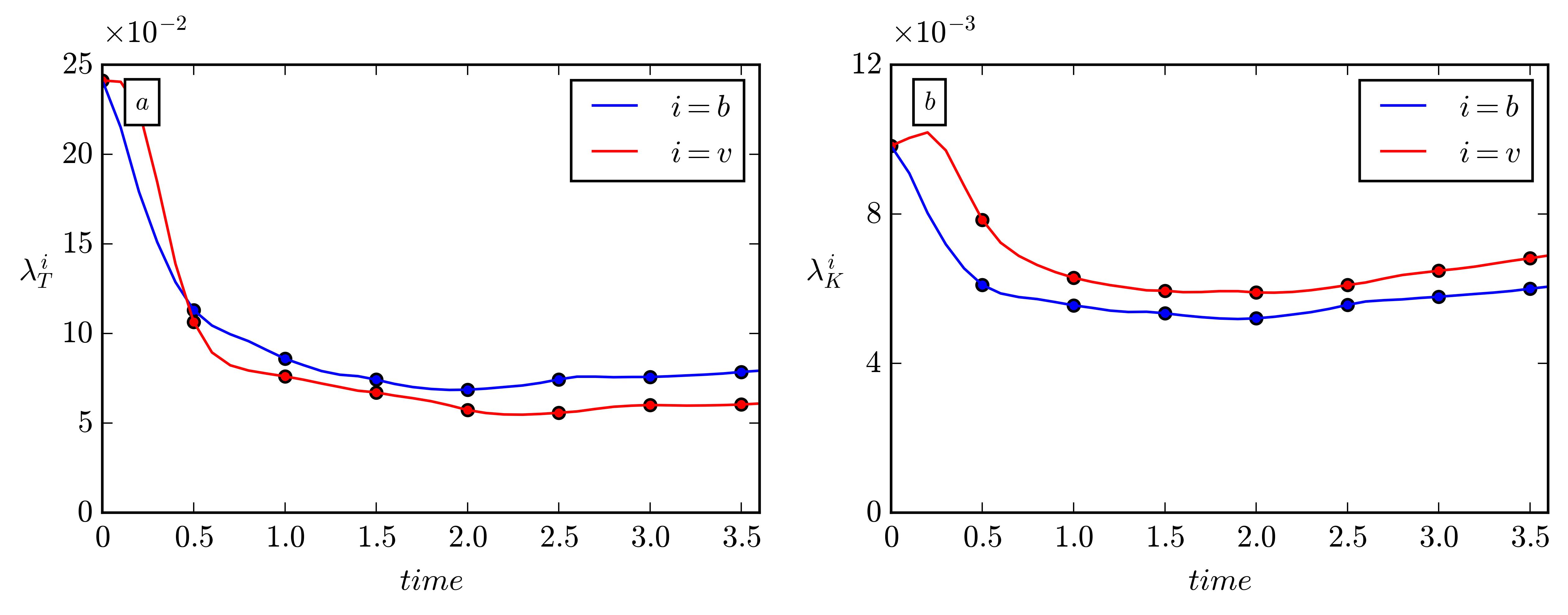}}
\caption{Characteristic length scales for the 2D MHD turbulent numerical simulation over time. (a) Magnetic ($ \lambda_T^{b} $) and kinematic ($ \lambda_T^{v} $) Taylor length scales. (b) Magnetic ($ \lambda_K^{b} $) and kinematic ($ \lambda_K^{v} $) Kolmogorov length scales. \label{fig: Fig1}}
\end{figure*}

The velocity field $ \boldsymbol{v}\left( \boldsymbol{x}, t \right) $ and the magnetic scalar potential $ a(\boldsymbol{x},t) $ are obtained by solving the set of 2D incompressible MHD equations (\ref{eq: AdvectionDiffusionVorticity})-(\ref{eq: AdvectionDiffusionMagnetic}) on a periodic domain $ \left[ x, y\right] \in \left[0, 2 \pi \right]^2 $ \citep{servidio2010statistics}. We compute the nonlinear terms using a pseudo-spectral technique, applying a 2/3 dealiasing rule. Time integration is achieved through a classical second order Runge-Kutta method with a meshgrid resolution of $ N = 2048 $ points. The spatial gradients are obtained through spectral differentiation. As a post-processing step, we apply a gaussian filter of size $ \sigma = 3 $ to the Eulerian and Lagrangian barrier fields. The kinematic viscosity $ \nu $ and the magnetic diffusivity $ \eta $ are both set to $ 4 \cdot 10^{-4} $. The numerical simulation spans a temporal domain of $ \left[0, 3.6 \right] $ and we record snapshots every $ \Delta t = 0.1 $. In total we have 37 snapshots that resolve the decaying 2D MHD turbulence simulation at high fidelity. We impose large scale random initial conditions, for both magnetic and velocity fields, in order to initiate a turbulent cascade and we randomly populate the modes $ 3 \leq k \leq 7 $. Our initial conditions mimic a large-to-small scale turbulence cascade, as observed in the solar wind, where energy is injected at the outer scales and cascades down to the smaller scales. Figure \ref{fig: Fig1} shows the Kolmogorov and Taylor length scales of the magnetic and velocity field as a function of time. \\ \\
In the following, we compare active magnetic, momentum and advective transport barriers at different times of the 2D MHD turbulence simulation. We first compute Eulerian barriers at time $ t= 2.0 $ and then extract Lagrangian barriers over time-interval $ t_0 = 1.0 $ and $ t_1 = 3.0 $. 
\subsubsection{Eulerian barriers}
\begin{figure*}[h]
\vskip1pc
\centering{\includegraphics[width=1\textwidth]{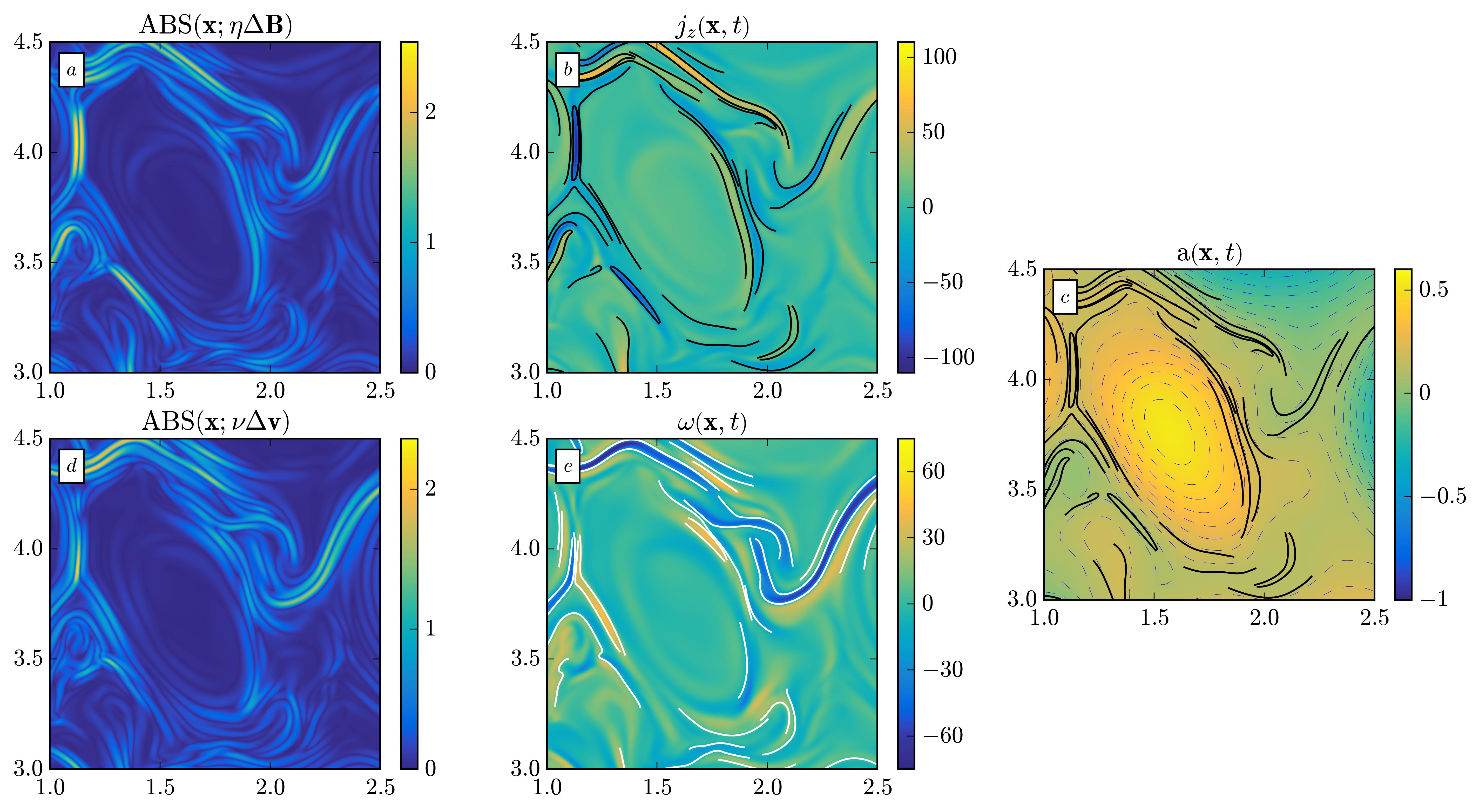}}
\caption{Comparison of instantaneous active magnetic (a,b) and momentum (d,e) barriers (f) over the domain $ \left[ 1, 4\right] \times \left[3, 6\right] $ of our 2D MHD turbulence example at time $ t = 2.0 $. Panel (c) displays the magnetic scalar potential $ a(\boldsymbol{x},t) $, with the dashed contours marking its level sets. The solid black and white curves respectively indicate the strongest active magnetic and momentum barriers.
\label{fig: Fig2}}
\end{figure*}

For the instantaneous magnetic and momentum barrier calculations, we use snapshots of the vorticity and electric current density field at fully developed turbulence at time $ t = 2.0 $ (see Fig. \ref{fig: Fig2}). In 2D incompressible MHD flows, both the magnetic and momentum barrier fields are Hamiltonian. Every level set of the electric current density $ j_z $ (panel b) is an exact barrier to the diffusive magnetic flux. Similarly, level sets of the vorticity $ \omega $ (panel e) qualify as perfect barriers to the diffusive momentum transport. Out of the infinitely many candidate curves, we seek the most influential momentum and magnetic barriers over the domain $ \left[1, 2.5\right] \times \left[3, 4.5\right] $ (see Fig. \ref{fig: Fig2}). We systematically extract these barriers by following the procedure outlined in Algorithm \ref{alg: InfluentialActiveBarrier2D}. For this purpose, we first compute the $ \mathrm{ABS} $ field associated to the instantaneous magnetic and momentum barrier fields (see panels a,d). We then launch streamlines of the corresponding barrier field from local maxima of the $ \mathrm{ABS} $ field to obtain exact active transport barriers. Here, we set $ \varepsilon_{\mathrm{ABS}} $ to be equal to the spatial average of the $ \mathrm{ABS} $ in the selected domain. Additionally, the minimum barrier length is set to $ \ell_{min} = \lambda^{i}_T $ and the minimal distance between two influential active barriers is $ \mathrm{d}_{min} = \lambda^{i}_{K} $. For momentum barriers (white curves in Fig. \ref{fig: Fig1}) we use the maximum kinematic length scales ($i=v$), whereas for the magnetic barriers (black curves in Fig. \ref{fig: Fig1}) we use the maximum magnetic length scales ($i=b$).
 \\ \\
In panel (c), we have included a snapshot of the magnetic potential $ a\left(\boldsymbol{x}, t\right)$, which is a frequently used coherent structure diagnostic in MHD flows \citep{servidio2010statistics, servidio2011magnetic}. Level sets of $ a\left(\boldsymbol{x}, t\right) $ correspond to magnetic field lines and are highlighted as grey dashed contours. The scalar magnetic potential shows an elliptic island that is surrounded by a complex pattern of active magnetic barriers. Specifically, these barriers separate elongated strips of intense electric current density, that are visible as ridges and trenches of $ j_z\left(\boldsymbol{x},t\right) $ \citep{cowley1997current, zhdankin2013statistical, donato2013identify}. These elongated peaks and troughs in the electric current density field manifest as electric current sheets, which play an important role in magnetic reconnection—a process where magnetic energy is converted to the kinetic and thermal energy of the particles \citep{biskamp1986magnetic, biskamp1994magnetic}. We observe a similar pattern in the vorticity field, where vorticity filaments are separated by momentum barriers (see white curves in Fig. \ref{fig: Fig2}).
\begin{figure*}[h]
\vskip1pc
\centering{\includegraphics[width=1\textwidth]{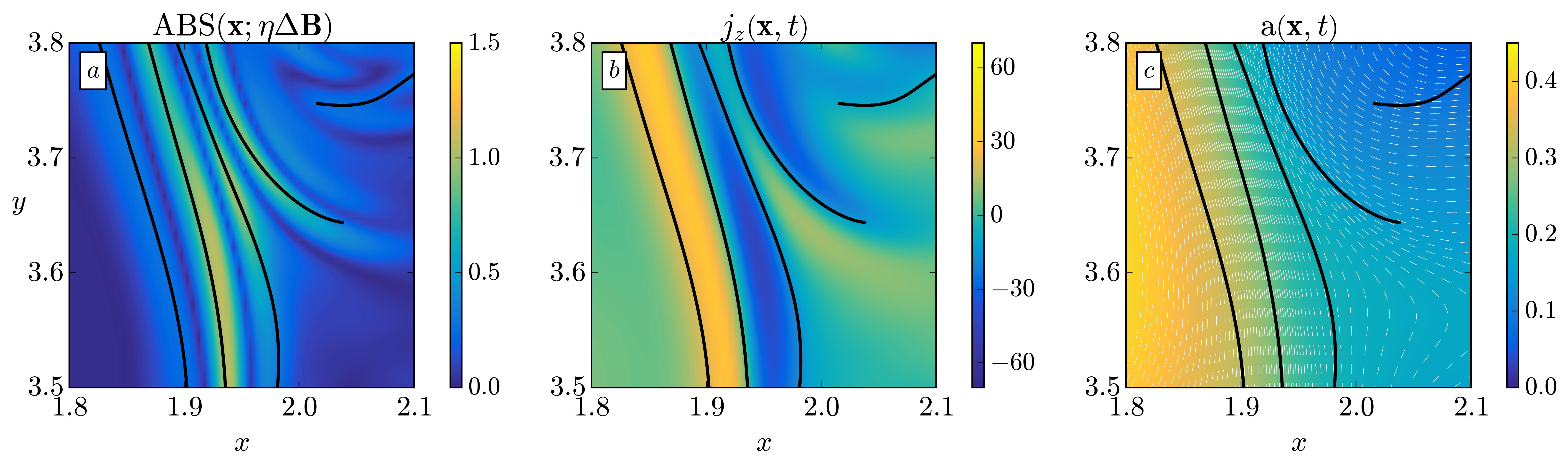}}
\caption{Comparison of active magnetic barriers (a) with the electric current density (b) and magnetic scalar potential (c) over the domain $ \left[ 1.5, 1.9\right] \times \left[3.7, 4.1\right] $.
\label{fig: Fig3}}
\end{figure*} \\  \\
Figure \ref{fig: Fig3} focuses on adjacent electric current sheets in the region $ \left[ 1.8, 2.1 \right] \times \left[ 3.5, 3.8 \right] $. We first emphasize that the active barriers closely align with ridges of the underlying $ \mathrm{ABS} $ field, as already suggested in section \ref{subsec: ActiveMagneticBarriers}. The principal electric current sheets are clearly visible as trenches and ridges of $ j_z\left(\boldsymbol{x}, t\right) $ which are surrounded by a set of active magnetic barriers (black curves). Active magnetic barriers provide a clear demarcation of electric current sheets, suggesting no diffusive transport of the magnetic field between adjacent current sheets. Instead, dissipation of the magnetic field is constrained to occur along the active magnetic barriers, as these barriers are defined as curves tangential to the diffusive term in the magnetic transport equation (\ref{eq: B_dot}). Despite providing critical information about underlying magnetic coherent structures, active magnetic barriers remain generally hidden in magnetic potential plots.
\subsubsection{Lagrangian barriers}
\begin{figure*}[h]
\vskip1pc
\centering{\includegraphics[width=1\textwidth]{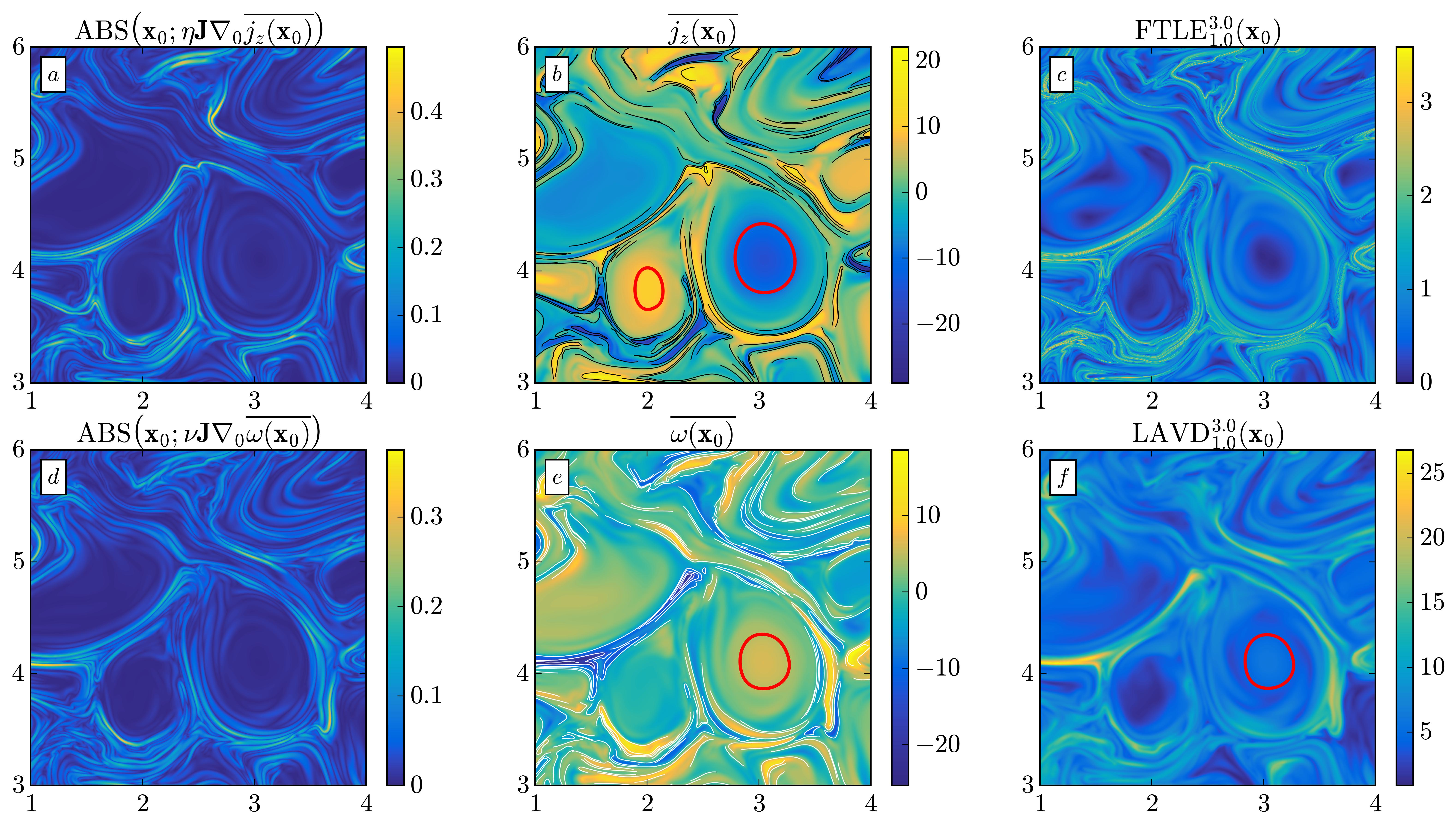}}
\caption{Comparison of Lagrangian active magnetic (a,b) and momentum (d, e) barriers with classic LCS diagnostics such as the $ \mathrm{FTLE}_{t_0}^{t_1} $ (c) and $ \mathrm{LAVD}_{t_0}^{t_1} $ (f) for our 2D MHD turbulence example over the time interval $ t_0 = 1.0 $ and $ t_1 = 3.0 $.
\label{fig: Fig4}}
\end{figure*}

For the Lagrangian barrier calculations, we first compute the Lagrangian averages $ \overline{j_z\left( \boldsymbol{x}_0\right)} $ and $ \overline{\omega\left( \boldsymbol{x}_0\right)} $ along fluid trajectories using all the available snapshots between $ t_0 = 1.0 $ and $ t_1 = 3.0 $ over the domain $ \left[1, 4\right] \times \left[3, 6\right] $. Based on that, we compute expressions for the active barrier fields from (\ref{eq: Theorem2_Lagrangian}) and (\ref{eq: 2D_Lagrangian_momentum}). To visualize advective LCSs at time $ t = 1.0 $, we plot the $ \mathrm{FTLE}_{t_0}^{t_1} $ and $ \mathrm{LAVD}_{t_0}^{t_1} $ fields over the initial conditions $ \boldsymbol{x}_0 $. We recall that ridges of the $ \mathrm{FTLE}_{t_0}^{t_1} $ field mark initial positions of the most repelling material lines (repelling LCS), whereas convex level sets of $ \mathrm{LAVD}_{t_0}^{t_1} $ surrounding an isolated local maximum indicate rotationally coherent structures (elliptic LCS). \\ \\
Figure \ref{fig: Fig4} shows the $ \mathrm{ABS} $ field computed for the Lagrangian magnetic and momentum barrier fields from (\ref{eq: Theorem2_Lagrangian}-\ref{eq: 2D_Lagrangian_momentum}) (see panels a, b). We then extract exact active material barriers by following the procedure outlined in Algorithm \ref{alg: InfluentialActiveBarrier2D}. For the active Lagrangian magnetic barrier calculations we set $ \ell_{min} = \lambda_T^{b} $ and $ \mathrm{d}_{min} = \lambda_K^{b} $, whereas for the momentum barriers we use the corresponding  kinematic length scales. Here, the black curves mark active magnetic material barriers, whereas the white curves indicate momentum blocking material barriers. Similarly to the case of the Eulerian barriers, ridges of the $ \mathrm{ABS} $ field closely align with exact active transport barriers in our Lagrangian computations (see Fig. \ref{fig: Fig4}). Active magnetic barriers (black curves in Fig. \ref{fig: Fig4}) mark sharp edges in the Lagrangian-averaged electric current density field $\overline{j(\boldsymbol{x}_0)}$, thereby separating the domain into areas with minimal time-averaged magnetic diffusion. Likewise, momentum barriers occur at sharp edges of the Lagrangian-averaged vorticity field $\overline{\omega(\boldsymbol{x}_0)}$. Note that, the Hamiltonians $ \overline{j(\boldsymbol{x}_0)} $ and $\overline{\omega(\boldsymbol{x}_0)}$ resemble features found in the $ \mathrm{FTLE}_{t_0}^{t_1} $ and $ \mathrm{LAVD}_{t_0}^{t_1} $ fields. This is to be expected because level sets of Lagrangian-averaged scalar fields occasionally relate to advective LCSs, that are obtained from purely kinematic computations \citep{kelley2013lagrangian, hadjighasem2017critical}. 
 \\ \\
Next, we compute LAVD-based vortex boundaries as outermost closed and convex level sets surrounding a unique local maximum of $ \mathrm{LAVD}_{t_0}^{t_1} $ \citep{Haller2016}. In practice, we relax the strict convexity requirement in order to allow for small initial filamentations that arise due to the finite grid of the numerical computations. The convexity deficiency is defined as
\begin{equation}
c_d = \frac{\left|A-A_{convex}\right|}{A},
\end{equation} where $ A $ is the area of the closed contour and $ A_{convex} $ is the area of its convex hull. We only retain large-scale vortices whose perimeter is greater than $ \pi\lambda_T^{v} $ and whose convexity deficiency is less than $ 10^{-6} $. In 2D incompressible MHD flows the Lagrangian active barrier fields are governed by the appropriate Hamiltonian function $ \mathcal{H} $. Therefore, we can extract elliptic active barriers as outermost convex level sets of $ \mathcal{H} $ surrounding a unique local maximum of $ \left|\mathcal{H}\right| $ \citep{Haller2020}. For active Lagrangian magnetic barriers we set $ \mathcal{H} = \overline{j(\boldsymbol{x}_0)}$, whereas for the momentum barriers we set $ \mathcal{H} = \overline{\omega(\boldsymbol{x}_0)}$.
The elliptic barriers computed from the underlying scalar field are shown as red curves in Fig. \ref{fig: Fig4}. The momentum and $ \mathrm{LAVD} $-based vortices are practically indistinguishable, whereas $ \overline{j_z\left(\boldsymbol{x}_0\right)} $ shows the existence of a magnetic vortex pair (see panel b). 
\begin{figure*}[h]
\vskip1pc
\centering{\includegraphics[width=1\textwidth]{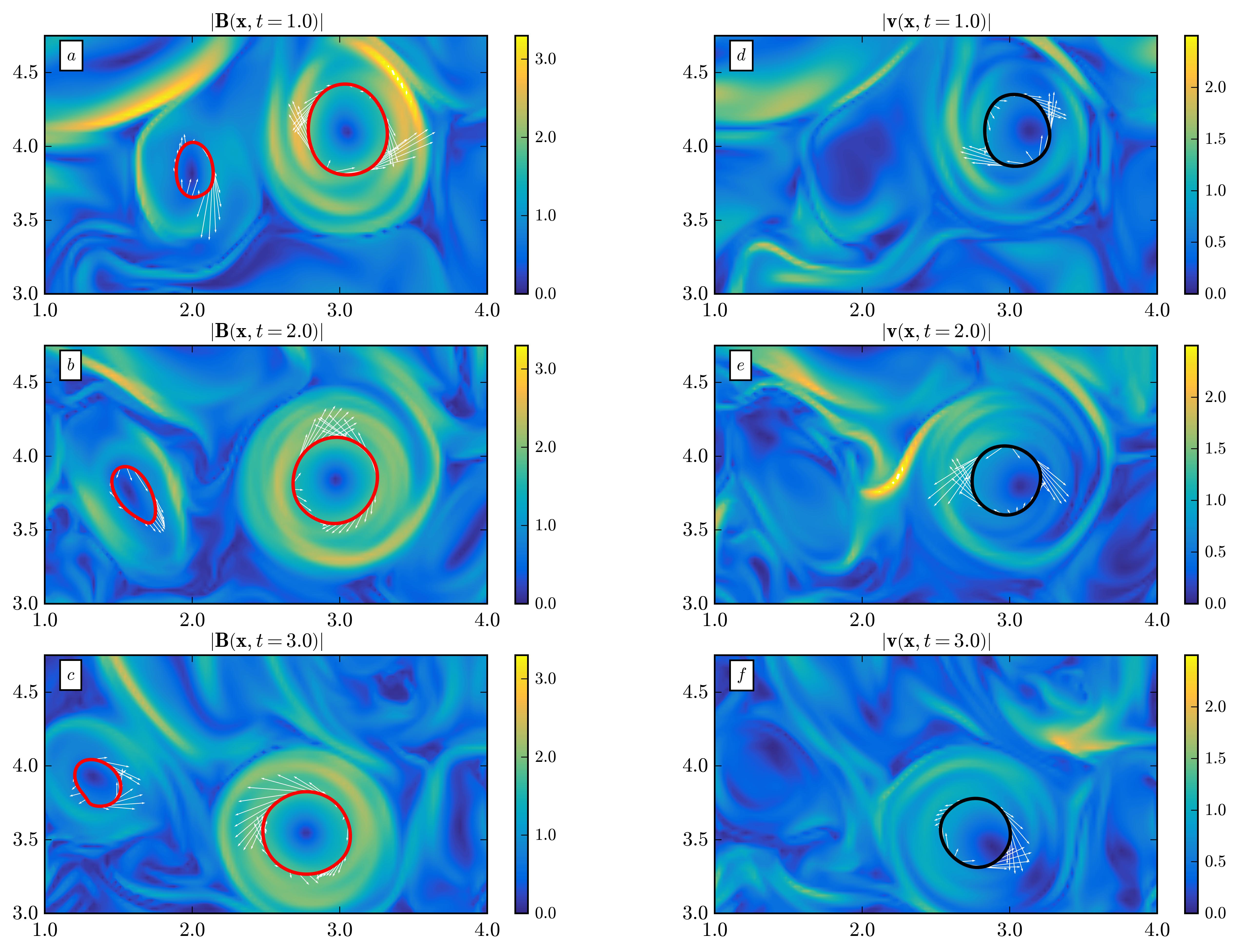}}
\caption{Evolution of active material elliptic barriers to the diffusive transport of the magnetic field (red curve in left column) and linear momentum (black curve in right column). The active vortices are superimposed on the distribution of the norm of the magnetic field (i.e. square-root of the magnetic energy) and linear momentum (normalized by $ \rho $). The vector fields in the left and right column respectively indicate the instantaneous active magnetic and momentum barrier fields at time $ t $. \label{fig: Fig5}}
\end{figure*} \\ \\
We now examine the temporal evolution of the magnetic vortex pair (depicted by red curves in the left column of Fig. \ref{fig: Fig5}) and illustrate its impact on the magnetic energy landscape within the time span $[1.0, 3.0]$. Additionally, we overlay the advected momentum-based vortex (represented by the black curve in the right column of Fig. \ref{fig: Fig5}) on top of the normalized linear momentum field. Notably, we observe that both active barriers exhibit no signs of filamentation throughout the entire duration. This is consistent with our earlier expectation for elliptic coherent structures. Furthermore, the active magnetic barriers keep enclosing regions of low magnetic field intensity throughout the extraction period. Similarly, active barriers consistently encapsulate small values of the linear momentum norm. \\ \\
Figure \ref{fig: Fig5} also shows the instantaneous active barrier fields along the respective active magnetic and momentum barriers. Note that the extracted barriers closely align with the underlying instantaneous active barrier fields for most of the time. Nonetheless, there are some notable exceptions, indicating that these barriers do not exactly minimize the instantaneous diffusive transport of the magnetic field or momentum at every time instance. Instead, active material barriers minimize the underlying diffusive transport of the magnetic field or momentum in a time-averaged sense.
\subsection{Three-dimensional MHD Turbulence}
\label{subsec: 3DMHD}
\begin{figure*}
\vskip1pc
\centering{\includegraphics[width=1\textwidth]{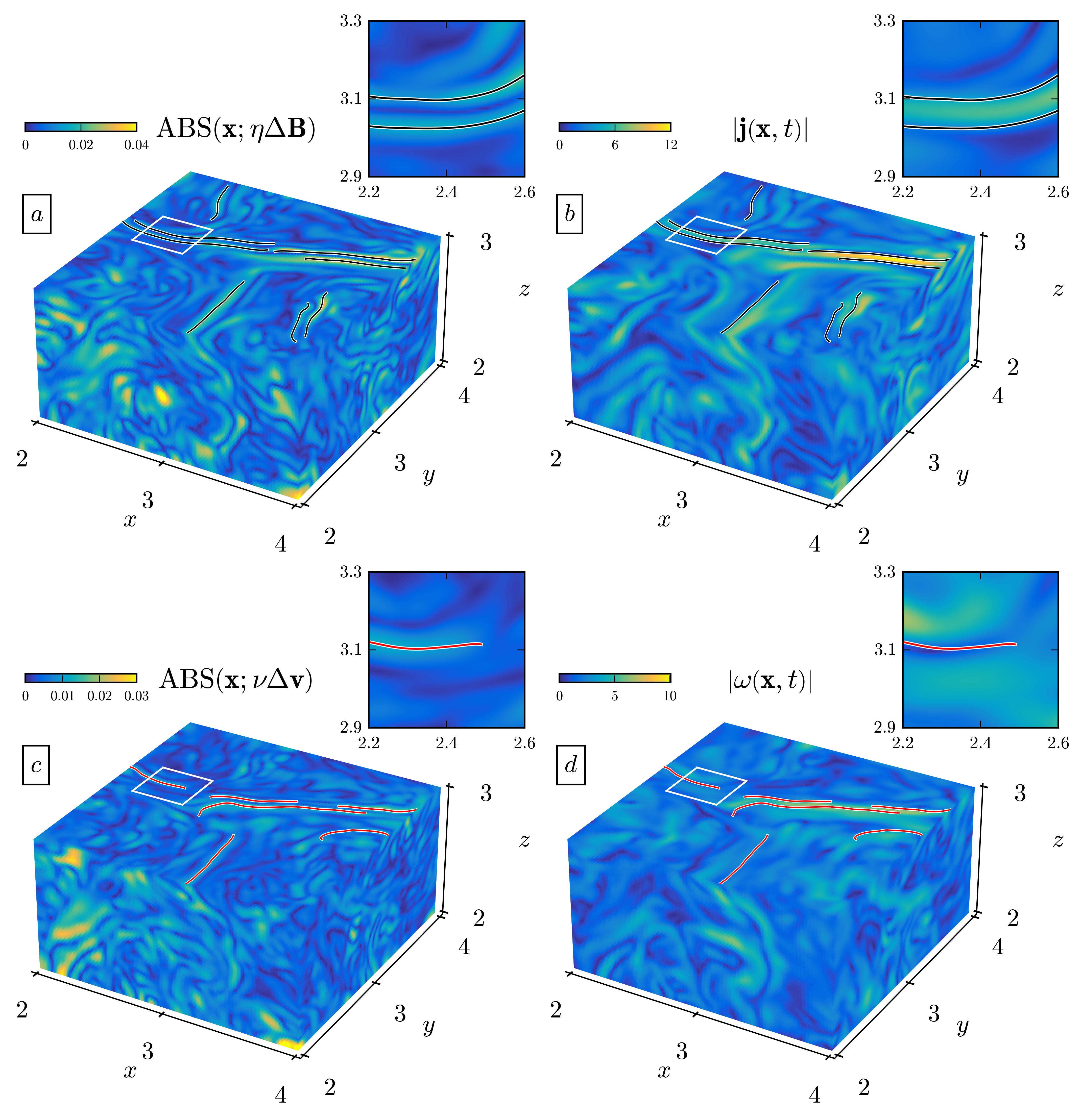}}
\caption{Comparison of Eulerian active magnetic and momentum barriers in the 3D MHD turbulence simulation from Johns Hopkins Turbulence Database (JHTDB) with the normed electric current density and the vorticity. The black and red curves are respectively the ridges of the magnetic and momentum-based $ \mathrm{ABS} $ fields on the $ z=3 $ plane. \label{fig: Fig6}} 
\end{figure*}

\begin{figure*}
\vskip1pc
\centering{\includegraphics[width=1\textwidth]{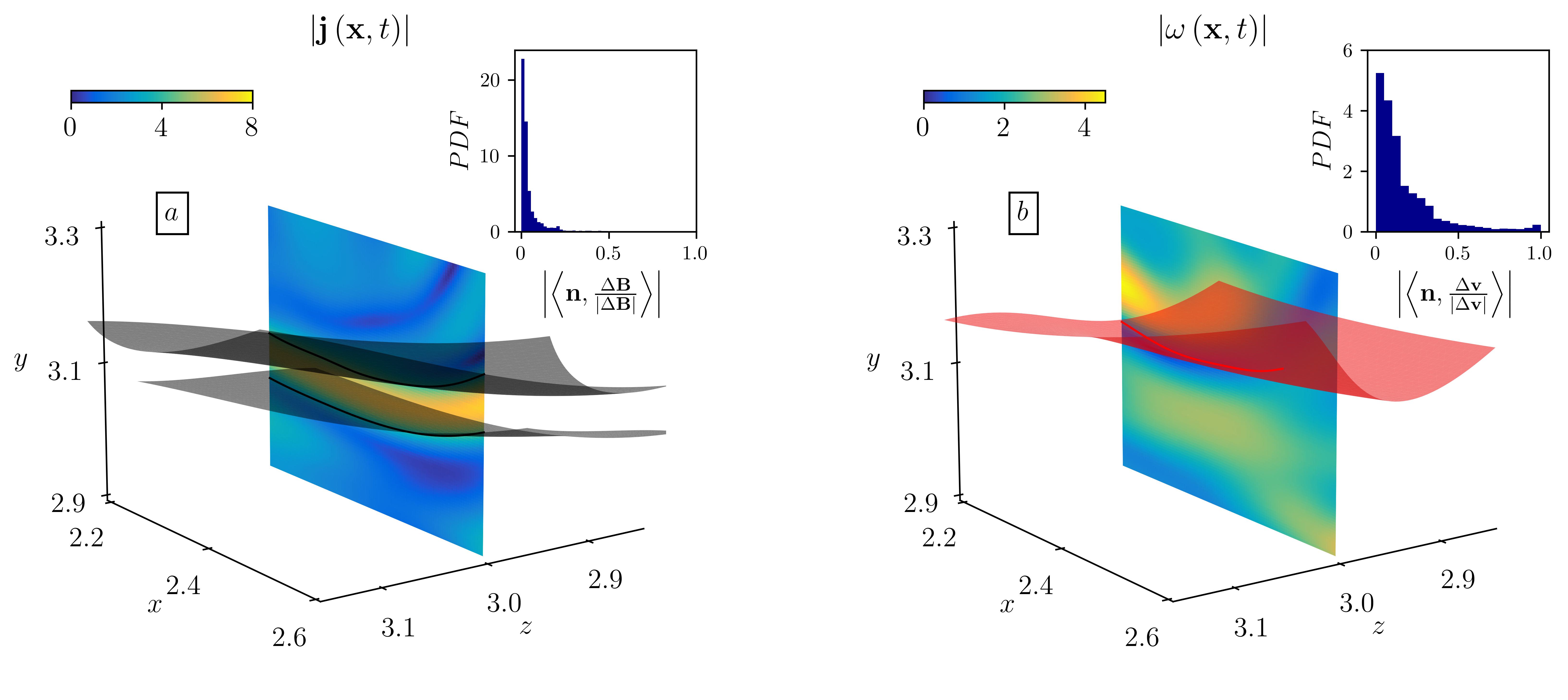}}
\caption{Active magnetic (a) and momentum (b) barriers in the zoomed inset from Fig. \ref{fig: Fig6} identified using Algorithm \ref{alg: InfluentialActiveBarrier3D} plotted on top of the normed electric current density (a) and vorticity (b). The insets show the probability distribution of the normed inner product between the normal vector of the surface and the underlying barrier fields. \label{fig: Fig7}}
\end{figure*}

In the following, we use forced MHD turbulence data from the Johns Hopkins Turbulence Database (JHTDB) \citep{perlman2007data, li2008public, aluie2009hydrodynamic, eyink2013flux}. The data was generated by a direct numerical simulation of the 3D incompressible MHD equations, in a cubic domain of size $ 2\pi $ with periodic boundary conditions and resolution $ 1024^3 $. The kinematic viscosity $ \nu $ and magnetic diffusivity $ \eta $ are both equal to $ 1.1 \cdot 10^{-4} $ and the kinematic and magnetic Kolmogorov length scales are respectively $ \lambda_K^{v}= 3.3 \cdot 10^{-3} $ and $ \lambda_K^{b}= 2.8 \cdot 10^{-3} $. The flow is forced at large scales in the $ x-y $ plane by a steady Taylor-Green body force
\begin{equation}
\boldsymbol{f} = f_0\left[ \sin(k_fx)\cos(k_fy)\cos(k_fz) \boldsymbol{e}_x - \cos(k_fx)\sin(k_fy)\cos(k_fz) \boldsymbol{e}_y\right],
\end{equation} with $ k_f = 2 $. \\ \\
In Fig. \ref{fig: Fig6} we compare the $ \mathrm{ABS} $ field for the magnetic (panel a) and momentum (panel c) barrier fields with the electric current density (panel b) and the normed vorticity (panel d) over the domain $ (x,y,z) \in \left[2,4\right] \times \left[2,4\right] \times \left[2,3\right] $. The zoomed inset in Fig. \ref{fig: Fig6}a shows two prominent ridges (black curves) of the $ \mathrm{ABS}\left(\boldsymbol{x};\eta \Delta \mathbf{B} \right) $ that delineate the boundary of an electric current sheet (see zoomed inset in panel b).  Similarly, ridges of the $ \mathrm{ABS}\left(\boldsymbol{x};\nu \Delta \boldsymbol{v} \right) $ (red curves) wrap around vorticity filaments also in 3D (see zoomed inset in panel d). \\ \\
Next, we focus on the region $ x, y \in \left[2.2, 2.6 \right] \times \left[2.9, 3.3 \right] $ in the $ z=3 $ plane which is shown in the zoomed inset of Fig. \ref{fig: Fig6} and extract instantaneous active magnetic and momentum barriers using the Algorithm \ref{alg: InfluentialActiveBarrier3D}. We approximate Eulerian active magnetic barriers, by fitting polynomial surfaces of degree $ \mathrm{d} = 3 $ to streamlines of (\ref{eq: Theorem1_Eulerian}) launched from ridges (black curves) of $ \mathrm{ABS}\left(\boldsymbol{x};\eta \Delta \mathbf{B} \right) $. Here, we set $ \varepsilon_{\mathrm{ABS}} $ to be equal to the spatial average of $ \mathrm{ABS} $ in the selected domain and the maximum arclength $ \ell_{max} $ of the streamline is chosen to be $ 100 \lambda_K^{b} $. For the momentum blocking barriers we follow a similar reasoning and replace the magnetic with the momentum quantities. Figure \ref{fig: Fig7}a displays two active magnetic barriers (black surfaces) that trace out the boundary of an electric current sheet. Similarly, the red surface in Fig. \ref{fig: Fig7}b corresponds to an approximate momentum-barrier separating two vorticity filaments. \\ \\
To test the transport blocking ability of the identified surface with respect to the underlying barrier field, we compute the pointwise normalized flux by taking the inner product between the normal vector $ \boldsymbol{n} $ of the surface and the corresponding normalized active barrier field. If the surfaces computed according to Algorithm \ref{alg: InfluentialActiveBarrier3D} were exact streamcurves of the barrier equation, then their normals would be pointwise perpendicular to the underlying barrier field. The inset in panel (a) displays the probability distribution of the normed inner product between $ \boldsymbol{n} $ and the unit vector of $ \Delta \mathbf{B} $ over both active magnetic barriers. Similarly, the inset of panel (b) shows the probability distribution of the pointwise tangency between the active momentum barrier field and the corresponding momentum blocking surface (red). Both distributions show a prominent peak at $ 0 $. This suggests that barriers obtained according to Algorithm \ref{alg: InfluentialActiveBarrier3D} are close approximations to perfect active barriers.

\section{Conclusion}
Appropriately modifying the recent active barrier theory of \cite{Haller2020} for Navier-Stokes flows, we have identified active magnetic barriers as material surfaces that minimize the diffusive magnetic flux in 2D and 3D MHD turbulence. These distinguished coherent structure boundaries, locally partition the domain into two regions with minimal diffusion of the magnetic field. We have also compared active magnetic barriers with linear momentum barriers and advective LCSs. Our analysis shows that active magnetic barriers provide objective barriers that inhibit magnetic diffusion. We have also devised an algorithm to extract the most influential active magnetic barriers in both 2D and 3D MHD flows. \\ \\ 
In 2D incompressible MHD, active magnetic barriers can directly be obtained as level curves of appropriate Hamiltonians that are a function of the electric current density. This stems from the fact that the equations governing these barriers form autonomous, planar Hamiltonian systems. Of the infinitely many barrier candidates, we have computed the most influential barriers as distinguished streamline segments of the appropriate active barrier field. These segments are launched from local maxima of the $ \mathrm{ABS} $ field (see Algorithm \ref{alg: InfluentialActiveBarrier2D}). 
\\ \\
A physical take-away message from our 2D MHD turbulence example is that the strongest Eulerian active magnetic barriers separate electric current sheets and hence induce zero magnetic diffusion across them. Instead, the diffusion of the magnetic field occurs along the active magnetic transport barriers we have identified. Secondly, we have computed active magnetic vortices as parametric curves from specific level sets of the Hamiltonian. As expected, the identified active magnetic vortices minimize the diffusive transport of the magnetic field and maintain coherence, i.e., they do not filament. Additionally, active Lagrangian magnetic vortices consistently encapsulate areas of low magnetic energy. Overall, our numerical computations show that active magnetic barriers generally differ from advective and linear momentum barriers. \\ \\
We have similarly obtained active magnetic barriers in 3D MHD turbulence over a 2D cross-section of the flow. Note, however, that in 3D the active barrier fields are no longer Hamiltonian and hence the electric current density does not directly appear in the active magnetic barrier equations. Interestingly, our numerical results on the Johns Hopkins MHD Turbulence Dataset suggest that the most influential active magnetic barriers in 3D arise at the interface between adjacent current sheets. Analogously to the case of the 2D MHD turbulence, this implies vanishing magnetic diffusion across nearby current sheets. \\ \\
The objective active magnetic barriers described here are intrinsic physical features of the fluid and contribute to the understanding and identification of various turbulent flow structures in MHD flows. Future research should investigate Lagrangian magnetic, momentum and advective transport barriers in the solar atmosphere, thereby expanding on the studies from \cite{nobrega2016cool, leenaarts2018tracing}. Additionally, we plan to relate the identified active magnetic barriers to dissipation of electromagnetic energy \citep{chian2023intensification, silva2024magnetohydrodynamic, finley2022stirring} and thermal coherent structures \citep{amari2015small, cranmer2019properties, loukitcheva2015millimeter} within the solar atmosphere.

\appendix

\section{Proof of Theorem \ref{th: Theorem2}}
\label{app: Theorem 2}
We recall that for 2D incompressible MHD flows, the magnetic field satisfies
\begin{equation}
\mathbf{B}\left( \boldsymbol{x},t \right) = -\mathbf{J}\nabla a \left( \boldsymbol{x},t \right), \label{eq: Hamiltonian_B_appendix}
\end{equation} with $ \mathbf{J} = \begin{pmatrix}
0 && -1 \\ 1 && 0
\end{pmatrix} $. The scalar magnetic potential $ a \left( \boldsymbol{x},t \right) $ satisfies the advection diffusion equation (\ref{eq: AdvectionDiffusionMagnetic}) and its solution along fluid trajectories $ \mathbf{F}_{t_0}^{t_1}\left( \boldsymbol{x}_0\right) $ is given by
\begin{align}
a \left(\mathbf{F}_{t_0}^{t_1}\left(\boldsymbol{x}_0\right),t\right)-  a (\boldsymbol{x}_0) &= \eta\int_{t_0}^{t_1} \Delta a\left(\mathbf{F}_{t_0}^{t}\left( \boldsymbol{x}_0\right), t \right) dt \nonumber \\
&= -\eta \int_{t_0}^{t_1} j_z\left(\mathbf{F}_{t_0}^{t}\left( \boldsymbol{x}_0\right), t \right) dt \nonumber \\
&= -\eta \left(t_1-t_0\right) \overline{j_z\left( \boldsymbol{x}_0, t\right)}, \label{eq: IntegralSolutionAdvectionDiffusion}
\end{align} where we have used the property $ j_z\left(\boldsymbol{x}, t \right) = -\Delta a\left(\boldsymbol{x}, t \right) $ and the temporal averaging operator (see formula (\ref{eq: Average})). With the notation 
\begin{equation}
\mathbf{B}_{t_0}^{t_1}\left(\boldsymbol{x}_0\right):=\mathbf{B}\left( \mathbf{F}_{t_0}^{t_1}\left( \boldsymbol{x}_0\right), t \right), \mathbf{B}_0\left(\boldsymbol{x}_0\right):=\mathbf{B}\left(\boldsymbol{x}_0,t_0 \right)
\end{equation} the integral form of the magnetic transport equation (\ref{eq: B_dot}) is
\begin{align}
\mathbf{B}_{t_0}^{t_1}\left(\boldsymbol{x}_0\right) &= \nabla \mathbf{F}_{t_0}^{t_1}\left( \boldsymbol{x}_0 \right)\mathbf{B}_0\left(\boldsymbol{x}_0\right)+\eta \int_{t_0}^{t_1}\nabla \mathbf{F}_{t}^{t_1}\left( \mathbf{F}_{t_0}^{t}\left( \boldsymbol{x}_0 \right) \right) \Delta \mathbf{B}\left( \mathbf{F}_{t_0}^{t}\left( \boldsymbol{x}_0 \right), t\right)dt \nonumber \\
&= \nabla \mathbf{F}_{t_0}^{t_1}\left( \boldsymbol{x}_0 \right)\mathbf{B}_0\left(\boldsymbol{x}_0\right) \nonumber \\
&+\eta \int_{t_0}^{t_1}\nabla \mathbf{F}_{t}^{t_1}\left( \mathbf{F}_{t_0}^{t}\left( \boldsymbol{x}_0 \right) \right) \nabla \mathbf{F}_{t_0}^{t}\left( \boldsymbol{x}_0 \right) \left[ \nabla \mathbf{F}_{t_0}^{t}\left( \boldsymbol{x}_0 \right) \right]^{-1} \Delta \mathbf{B}\left( \mathbf{F}_{t_0}^{t}\left( \boldsymbol{x}_0 \right), t\right)dt \nonumber \\
&= \nabla \mathbf{F}_{t_0}^{t_1}\left( \boldsymbol{x}_0 \right)\mathbf{B}_0\left(\boldsymbol{x}_0\right) +\eta \nabla \mathbf{F}_{t_0}^{t_1}\left( \boldsymbol{x}_0 \right) \int_{t_0}^{t_1}  \left[ \nabla \mathbf{F}_{t_0}^{t}\left( \boldsymbol{x}_0 \right) \right]^{-1} \Delta \mathbf{B}\left( \mathbf{F}_{t_0}^{t}\left( \boldsymbol{x}_0 \right), t\right)dt \nonumber \\
&= \nabla \mathbf{F}_{t_0}^{t_1}\left(\boldsymbol{x}_0\right) \left[\mathbf{B}_0\left(\boldsymbol{x}_0\right) +\eta \int_{t_0}^{t_1}  \left[ \nabla \mathbf{F}_{t_0}^{t}\left( \boldsymbol{x}_0 \right) \right]^{-1} \Delta \mathbf{B}\left( \mathbf{F}_{t_0}^{t}\left( \boldsymbol{x}_0 \right), t\right)dt \right] \nonumber \\
&= \nabla \mathbf{F}_{t_0}^{t_1}\left(\boldsymbol{x}_0\right) \left[\mathbf{B}_0\left(\boldsymbol{x}_0\right) +\eta \left(t_1-t_0\right) \overline{\left( \mathbf{F}_{t_0}^{t} \right)^{*} \Delta \mathbf{B}\left( \boldsymbol{x}_0\right)}\right] \nonumber \\
&= \nabla \mathbf{F}_{t_0}^{t_1}\left(\boldsymbol{x}_0\right) \left[\mathbf{B}_0\left(\boldsymbol{x}_0\right) + \left(t_1-t_0\right) \mathbf{\overline{b}}_{t_0, dif}^{t_1}\left( \boldsymbol{x}_0\right)\right], \label{eq: B_transport}
\end{align} where we used the property $ \nabla \mathbf{F}_{t_0}^{t_1}\left(\boldsymbol{x}_0\right) = \nabla \mathbf{F}_{t}^{t_1}\left(\mathbf{F}_{t_0}^{t}\left(\boldsymbol{x}_0\right)\right) \nabla \mathbf{F}_{t_0}^{t}\left(\boldsymbol{x}_0\right) $. By rearranging Eq.(\ref{eq: B_transport}), we then obtain
\begin{equation}
\frac{1}{t_1-t_0} \left[
\left( \mathbf{F}_{t_0}^{t_1} \right) ^{*} \mathbf{B}_{0} (\boldsymbol{x}_0)-\mathbf{B}_{0} (\boldsymbol{x}_0) \right] = \boldsymbol{\overline{b}}_{t_0, dif}^{t_1}\left( \boldsymbol{x}_0\right). \label{eq: B_transport_rearranged}
\end{equation} 
Inserting formula (\ref{eq: Hamiltonian_B_appendix})
into (\ref{eq: B_transport_rearranged}) yields
\begin{align}
\frac{1}{t_1-t_0} \left[
\left( \mathbf{F}_{t_0}^{t_1} \right) ^{*} \mathbf{B}_{0} (\boldsymbol{x}_0)-\mathbf{B}_{0} (\boldsymbol{x}_0) \right]  &= \frac{-1}{t_1-t_0}\left[\left( \mathbf{F}_{t_0}^{t_1} \right) ^{*} \mathbf{J}\nabla a (\boldsymbol{x}_0) -\mathbf{J}\nabla a_0 (\boldsymbol{x}_0) \right] \nonumber \\
&= \frac{-1}{t_1-t_0}\left[\left[ \nabla \mathbf{F}_{t_0}^{t_1} (\boldsymbol{x}_0) \right]^{-1} \mathbf{J}\nabla a \left(\mathbf{F}_{t_0}^{t_1}\left(\boldsymbol{x}_0\right),t\right) -\mathbf{J}\nabla_0 a (\boldsymbol{x}_0) \right], \label{eq: B2}
\end{align} where $ a\left( \boldsymbol{x}_0 \right):=a \left( \boldsymbol{x}_0,t_0 \right) $. Using now the chain rule for $ \nabla a \left( \mathbf{F}_{t_0}^{t_1}\left( \boldsymbol{x}_0 \right),t \right) $, we can write
\begin{align}
\left[ \nabla \mathbf{F}_{t_0}^{t_1} (\boldsymbol{x}_0) \right]^{-1} \mathbf{J}\nabla a \left(\mathbf{F}_{t_0}^{t_1}\left(\boldsymbol{x}_0\right),t\right) &= \left[ \nabla \mathbf{F}_{t_0}^{t_1} (\boldsymbol{x}_0) \right]^{-1} \mathbf{J} \left[ \nabla \mathbf{F}_{t_0}^{t_1} (\boldsymbol{x}_0) \right]^{-T}\nabla_0 a \left(\mathbf{F}_{t_0}^{t_1}\left(\boldsymbol{x}_0\right),t\right) \nonumber \\
&= \mathbf{J} \nabla_0 a \left(\mathbf{F}_{t_0}^{t_1}\left(\boldsymbol{x}_0\right),t\right) \label{eq: B3},
\end{align}
where we have used the fact that in 2D incompressible flows
\begin{equation}
\begin{pmatrix} a && b \\ c && d \end{pmatrix} \begin{pmatrix} 0 && -1 \\ 1 && 0 \end{pmatrix} \begin{pmatrix} a && c \\ b && d \end{pmatrix} = \begin{pmatrix} 0 && bc-ad \\ ad-bc && 0 \end{pmatrix} = \begin{pmatrix} 0 && -1 \\ 1 && 0 \end{pmatrix},
\end{equation} where $ \left[\nabla \mathbf{F}_{t_0}^{t_1} (\boldsymbol{x}_0)\right]^{-1} = \begin{pmatrix} a && b \\ c && d \end{pmatrix} $ and $ \mathrm{det}\left( \left[\nabla \mathbf{F}_{t_0}^{t_1} (\boldsymbol{x}_0)\right]^{-1} \right) = ad-bc = 1 $. By combining the relationships (\ref{eq: B2}-\ref{eq: B3}) we then obtain
\begin{align}
\frac{1}{t_1-t_0}\left[\left( \mathbf{F}_{t_0}^{t_1} \right) ^{*} \mathbf{B}_{0} (\boldsymbol{x}_0)-\mathbf{B}_{0} (\boldsymbol{x}_0) \right]&= \frac{-1}{t_1-t_0}\left[ \mathbf{J} \nabla_0 a \left(\mathbf{F}_{t_0}^{t_1}\left(\boldsymbol{x}_0\right),t\right)- \mathbf{J}\nabla_0 a (\boldsymbol{x}_0) \right] \nonumber \\
&= -\mathbf{J} \nabla_0 \left[a \left(\mathbf{F}_{t_0}^{t_1}\left(\boldsymbol{x}_0\right),t\right)- a (\boldsymbol{x}_0)\right] \label{eq: B5}.
\end{align} Finally, combining Eq. (\ref{eq: IntegralSolutionAdvectionDiffusion}) with Eqs.(\ref{eq: B5}) and (\ref{eq: B_transport_rearranged}) yields
\begin{align}
\boldsymbol{\overline{b}}_{t_0,dif}^{t_1}(\boldsymbol{x}_0) = \eta \mathbf{J} \mathbf{\nabla}_0 \overline{j_z\left( \boldsymbol{x}_0, t\right)}. \label{eq: A10}
\end{align} We then obtain the Eulerian counterpart by taking the infinitesimal limit of Eq.(\ref{eq: A10})
\begin{equation}
\boldsymbol{b}_{dif}(\boldsymbol{x},t) = \lim\limits _{t_1\rightarrow t_0=t}\boldsymbol{\overline{b}}_{t_0,dif}^{t_1}(\boldsymbol{x}_0) =  \eta \mathbf{J} \mathbf{\nabla} j_z\left( \boldsymbol{x}, t\right). \label{eq: A11}
\end{equation} This concludes the proof.



\bibliographystyle{elsarticle-num-names} 
\bibliography{references}


\end{document}